%% file: WSN_Journal_Paper.tex
\documentclass[11pt,draftcls,onecolumn]{IEEEtran}
\usepackage{amsmath,epsfig,subfigure}

\input{macros}

\newtheorem{proposition}{Proposition}
\newtheorem{definition}{Definition}

\title{Transform-based Distributed Data Gathering}
\author{Godwin~Shen,~\IEEEmembership{Student Member,~IEEE,}
        and~Antonio~Ortega,~\IEEEmembership{Fellow,~IEEE}
\thanks{Manuscript received September 24, 2009. This work was supported 
in part by NASA under grant AIST-05-0081.}
\thanks{G. Shen is with the Department
of Electrical Engineering, University of Southern California, Los Angeles,
CA, 90089 USA (e-mail: godwinsh@usc.edu).}% <-this % stops a space
\thanks{A. Ortega is with the Department
of Electrical Engineering, University of Southern California, Los Angeles,
CA, 90089 USA (e-mail: ortega@sipi.usc.edu).}% <-this % stops a space
}
\begin{document}
%\ninept
%
\maketitle
\begin{abstract}
A general class of unidirectional transforms 
is presented that can be computed in a distributed manner 
along an arbitrary routing tree. 
Additionally, we provide a set of conditions under 
which these transforms are invertible. 
These transforms can be computed as data is routed towards the 
collection (or sink) node in the tree and 
exploit data correlation between nodes in the 
tree. Moreover, when used in wireless sensor networks, 
these transforms can also leverage data received at 
nodes via broadcast wireless communications. 
Various constructions of unidirectional 
transforms are also provided for use in 
data gathering in wireless sensor networks. 
New wavelet transforms are also proposed which 
provide significant improvements 
over existing unidirectional transforms.
\end{abstract}

\begin{center} \bfseries EDICS Category: SEN-DIST \end{center}
%\begin{center} EDICS Category: SEN-DIST \end{center}

\section{Introduction}
\label{sec:intro}

In networks such as wireless sensor networks (WSNs), one major 
challenge is to gather data from a set of nodes and transfer it to a collection (or sink) node 
as efficiently as possible. 
Efficiency can be measured in terms of bandwidth utilization, energy consumption, etc. 
We refer to this as the \emph{data gathering problem}. 
The gathering is typically done in data gathering rounds or \textit{epochs} along a collection of routing paths to the sink, i.e., 
in every epoch each node forwards data that it has measured along a multi-hop path to the sink.
A simple gathering strategy is to have each node route raw data to the sink in a way that minimizes some cost metric, 
e.g., number of hops to the sink, energy consumption. This minimizes the amount of resources nodes use to transfer 
raw data to the sink and is the basis for many practical systems used in WSN 
such as the Collection Tree Protocol (CTP)~\cite{ctp}.
However, it has been recognized in the literature~\cite{Chong,broadcast2} that, in a WSN, 
(i) spatial data correlation may exist across neighboring nodes 
and (ii) nodes that are not adjacent to each other in a routing path can still 
communicate due to broadcasted wireless 
transmissions\footnote{Data transmissions in a WSN are typically broadcast~\cite{Cidon,broadcast1}, 
so multiple nodes can receive a single data transmission.}. 
%GSrev1 Rewrote for brevity
Raw data forwarding does not make use of these two facts, 
thus, it will not be the most efficient data gathering method in general.
%Raw data forwarding does not exploit spatial data correlation, nor does it use additional data 
%communication links induced by broadcast, thus, it will not be the most efficient data gathering method in general.

When spatial data correlation exists, it may be useful to apply \emph{in-network compression} distributed across the nodes 
to reduce this data redundancy~\cite{Chong}. More specifically, nodes can exchange data with their neighbors in order to 
remove spatial data correlation. This will lead to \emph{a representation requiring fewer bits per measurement 
as compared to a raw data representation}, 
also leading to reduced energy consumption, bandwidth usage, delay, etc. 
Since nodes in a WSN are severely energy-constrained~\cite{Chong,broadcast2,Wang}, some form of in-network processing 
that removes data redundancy will help reduce the amount of energy nodes consume in transmitting data to the sink. 
In this way the lifetime of a WSN can be extended. 
This could also be useful in bandwidth-limited applications~\cite{proakis,mechitov}. 
%This could also be useful in other bandwidth-limited applications such as underwater acoustic 
%networks~\cite{proakis} and structural health monitoring~\cite{mechitov}. 

Generally speaking, distributed spatial compression schemes 
require some form of data exchange between nodes. Therefore, one needs to select both 
\emph{a routing strategy} and \emph{a processing strategy}. 
The routing strategy defines what data communications nodes need to make 
and the processing strategy defines how each node processes data. 
There are a variety of approaches available, e.g., 
distributed source coding (DSC) techniques~\cite{slepian-wolf,pradhan}, 
%distributed source coding (DSC) techniques such as Networked Slepian-Wolf coding~\cite{slepian-wolf} and DISCUS~\cite{pradhan}, 
%distributed source coding (DSC) techniques such as Networked Slepian-Wolf coding~\cite{slepian-wolf}, 
transform-based methods like Distributed KLT~\cite{Gastpar}, Ken~\cite{ken}, PAQ~\cite{paq}, and 
wavelet-based approaches~\cite{Wagner1,Wagner2,Lozano,CiancioO:04,CiancioPOK:06,Ciancio:06,Shen1}. 
Note that DSC techniques do not require nodes to exchange data in order to 
achieve compression. Instead, each node can compress its own data using some statistical 
correlation model. Note, however, that an estimate of these models must be known at every node, so nodes will still 
need to do some initial data exchange in order to learn the models (after which compression can be 
done independently at each node). 
Our work only considers transform-based methods, which use linear transforms 
to decorrelate data while distributing transform computations across different nodes. 
%to de-correlate data while distributing transform computations across different nodes in the network. 
While we do not consider DSC approaches, our algorithms could be useful in the training phase of these methods to estimate correlation. 
Ken and PAQ are examples of approaches we consider, where data at each node is 
predicted using a linear combination of measurements from the node and 
measurements received from its neighbors. Similarly, the Distributed KLT, wavelet-based methods 
and many other related methods also use linear transforms to decorrelate data. 
Therefore, we can restrict ourselves to linear in-network transforms while still encompassing a general 
class of techniques.

Many of the existing transform-based methods \emph{propose a specific 
transform first}, then 
design routing and processing strategies that allow the transform to 
be computed in the network. Some examples are the wavelet transforms proposed in~\cite{Wagner1,Wagner2,CiancioPOK:06}, 
the Distributed KLT, Ken and PAQ. While these methods are good from a data decorrelation standpoint, 
\emph{the routing and processing strategies that are used to facilitate distributed processing may not always 
be efficient in terms of data transport cost}. In particular, nodes may have to transmit their own data 
multiple times~\cite{Wagner1,Wagner2}, nodes may need to transmit multiple 
%multiple times~\cite{Wagner1,Wagner2}, nodes may need to transmit multiple (or redundant) 
copies of the same coefficients~\cite{CiancioPOK:06}, 
or nodes may even need to transmit data away from the sink~\cite{Gastpar,Wagner1,Wagner2}. 
%As a representative example, \figurename~\ref{fig:BidirectionalExample} shows how to compute the transform in~\cite{Wagner1,Wagner2} 
%GSrev1 This example doesn't add much to the discussion. Removing for the sake of space.
%As a representative example, Fig.~\ref{fig:BidirectionalExample} shows how to compute the transform in~\cite{Wagner1,Wagner2} 
%in a distributed manner. Nodes are first divided into disjoint sets of even and odd nodes. 
%Initially nodes in the even set transmit raw data to their neighbors in the odd set, 
%then nodes in the odd set compute transform coefficients and transmit these coefficients back to 
%their even neighbors (and to the sink). Finally, the even nodes use these coefficients to compute their own coefficients, then they 
%route them to the sink. This forces many of the even nodes to transmit data away from the sink and also increases the 
%total communication cost of even nodes since each one must transmit its own data twice. 
As discussed in~\cite{Wagner2}, this sort of strategy can outperform raw data gathering for very dense networks, but 
it can lead to significant communication overhead for small to medium sized ones. 
%As discussed in~\cite{Wagner2}, this sort of strategy can outperform raw data gathering for relatively dense networks, but 
%it can lead to significant communication overhead for small to medium sized ones (less than 200 nodes). 
%Other related methods may also suffer from such inefficiencies.

%GSrev1 Removed to save on space
%\begin{figure}[htb]
% \centering
% \psfig{figure=BidirectionalExample.eps,width=12cm,height=5cm}
%% \centerline{\psfig{figure=BidirectionalExample.eps,width=12cm,height=5cm}}
%  \caption{{\footnotesize Example of bidirectional transform in~\cite{Wagner1,Wagner2}.
%  ``Even'' nodes 1, 3, 5, 7, 9 and 11 first transmit data to ``odd'' neighbors (left graph), 
%  then ``odd'' nodes 2, 4, 6, 8, 10, 12 and 13 use this data to compute detail coefficients. 
%  ``Odd'' nodes then transmit detail coefficients to ``even'' neighbors (center graph), then ``even'' nodes compute smooth coefficients
%  using these details.}}
%\label{fig:BidirectionalExample}
%\end{figure}

The results of our previous work~\cite{Shen1,Shen2} and of~\cite{Wagner2} 
demonstrate why transport costs cannot be ignored. 
One simple way to work around these issues is to first design an efficient routing tree (e.g., a shortest path 
routing tree, or SPT), then allow the transform computations 
to occur only along the routing paths in the tree. We call these types of schemes \emph{en-route in-network transforms}. 
These transforms (e.g., the wavelet transforms in~\cite{Lozano,CiancioO:04,CiancioPOK:06,Ciancio:06,Shen1}) 
will typically be more efficient since they 
are computed as data is routed to the sink along efficient routing paths. 
In addition to overall efficiency, these transforms can be easily integrated on top of existing routing protocols, i.e., 
a routing tree can be given by a protocol, then the transform can be constructed along the tree. 
This allows such schemes to be easily usable in a WSN - as demonstrated by the SenZip~\cite{senzip} compression 
tool, which includes an implementation of our algorithm in~\cite{Shen1} - as well as other types of data gathering 
networks~\cite{proakis,mechitov}. 

\begin{figure}[htb]
 %\centerline{\psfig{figure=BroadcastExample.eps,width=12cm,height=6cm}}
 %\centerline{\psfig{figure=BroadcastExample.eps,width=11cm,height=5.5cm}}
 \centerline{\psfig{figure=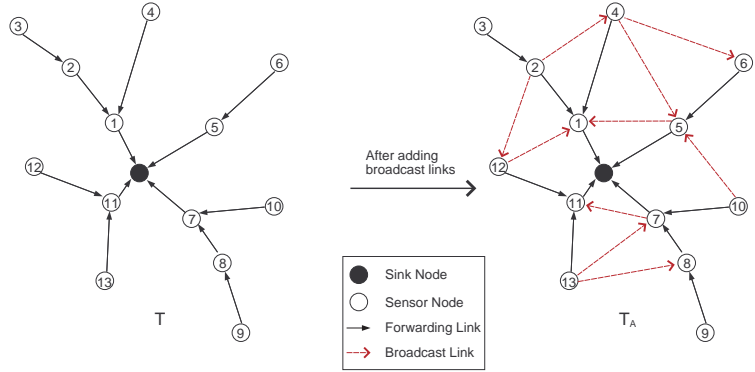,width=10cm,height=5cm}}
  \caption{{\footnotesize Example of routing tree and a tree augmented with broadcasts. 
  Solid arrows denote forwarding links along the tree and dashed arrows denote broadcast links.}}
\label{fig:BroadcastExample}
\end{figure}

We note that all existing en-route transforms 
start from well-known transforms, then modify them to work on routing trees. 
%Thus, they do not cover the full range of such transforms. 
%GSrev1 This doesn't flow too well. I rewrote it above.
%While existing en-route in-network transforms do lead to efficient data gathering, 
%they do not cover the full range of such transforms. 
%GSrev1 This sentence isn't relevant until later on
%Moreover, these transforms may not provide the most efficient representation of the underlying data. 
Instead, in this work we start from a routing tree $T$ and additional links given by broadcast 
(e.g., Fig.~\ref{fig:BroadcastExample}). 
We then pose the following questions: 
(i) what is the full set of transforms that can be computed as data is routed towards the sink 
along $T$ and (ii) what are conditions for invertibility of these transforms? 
\emph{The main goal of this work is to determine this general set of invertible, en-route in-network transforms}. 
%GSrev2 Added discussion here about quantization and encoding.
%AOrev2 minor editing 
%Also the point underlying the comments, was whether a specific transform could turn out to be ok in the lossless case, but somehow be bad for a lossy setting 
Note that in many transform-based compression systems, design or selection of a transform is considered separately from the design of a quantization and encoding strategy. This is done in practice 
in order to simplify the system design (e.g.,~\cite{Taubman}). 
In general 
certain properties of the transform (energy compaction, orthogonality) can serve as indicators of achievable performance in the lossy case. 
We adopt a similar approach in our work, 
choosing to only focus on the transform design. 
Simple quantization and encoding schemes can then be applied to the transform coefficients, 
as demonstrated in our experimental results. 
Joint optimization of routing and compression is also possible, as in~\cite{Pattem,Rickenbach} 
and our previous work~\cite{Shen2}, but this is beyond the scope of this work. 

In order to formulate this problem, we first note that the data gathering process consists of 
data measurement at each node and routing of data to the sink along $T$ done in 
accordance with some transmission scheduling, i.e., nodes transmit data along $T$ in a certain order. 
Also note that data is only transmitted along $T$ in the direction of the 
sink, i.e., data transmissions are \emph{unidirectional} towards the sink.
Moreover, each node can only process its own 
data with data received from other nodes that transmit before it, i.e., 
processing of data must be \emph{causal} in accordance with the transmission schedule. 
In particular, before each node transmits it will only have access 
to data received from nodes that use it as a relay in a 
multi-hop path to the sink (i.e., ``descendants'') and nodes whose data it receives but is not responsible 
for forwarding to the sink (i.e., ``broadcast'' neighbors). 
Whenever broadcast is used, data from a single node will often be available at multiple nodes. While 
this can help to decorrelate data even further (since more data will be available for transform computations at 
each node), it would be undesirable to transmit this same piece of data through multiple paths since this would 
increase the overall communication cost. 
Thus, in addition to causality and unidirectionality, the transform should also be \emph{critically sampled}, i.e., 
%AOrev1 Below was ambiguous
%there is exactly one transmission per transform coefficient. 
the number of transform coefficients that are computed and routed to the 
sink is equal to the number of nodes in the network. 
%This prevents nodes from forwarding redundant transform coefficients to the sink through multiple paths. 
We refer to causal, critically-sampled transforms that are computed in a unidirectional manner as \emph{unidirectional transforms}. 

As we will show, unidirectional transforms can be defined in terms of the routing tree, the broadcast links 
induced by the routing and the transmission schedule. Thus, given a tree and transmission schedule, the main problem 
we address in this work is to determine a set of necessary and sufficient conditions under which 
an arbitrary unidirectional transform is invertible. 
%We call this the \emph{invertible en-route transform problem}.
%GSrev1 Pulled this text from the following paragraphs and edited it as to highlight our contributions.
While unidirectional transforms have been proposed, to 
the best of our knowledge, none of the existing works have attempted to define the most general set of unidirectional transforms, 
nor has any attempt been made to find conditions under which such transforms are invertible. 
%GSrev1 Moved this statement here to help highlight our contribution
Our proposed theory also incorporates the use of broadcast data in a general setting. 
This leads us to develop transforms that use broadcasts in a manner not previously considered. 
This contribution is discussed in detail in Section~\ref{sec:in-network-processing}.

%GSrev1 Reworded this section a bit to highlight how our proposed transforms are different from existing ones.
%%GSrev1 OLD TEXT
%%In the context of WSNs, early work in~\cite{Lozano,CiancioO:04,CiancioPOK:06,Ciancio:06} developed 
%%en-route in-network wavelet transforms on 1D routing paths in WSNs. Extensions to 2D routing paths on arbitrary 
%%routing trees were made in~\cite{Shen1,Shen2} and extensions to graphs were proposed in~\cite{Shen4}.
%%The superiority of these 1D~\cite{CiancioPOK:06} and 2D~\cite{Shen1} en-route transforms 
%%over the method in~\cite{Wagner2} - which requires a great deal of backward communication - 
%%was demonstrated in~\cite{Shen1} in that the amount of energy consumption required to achieve a fixed 
%%reconstruction quality of the data is lower than the transform of~\cite{Wagner2}. 
%%GSrev1 NEW TEXT
In the context of wavelet transforms for WSNs, early work~\cite{Lozano,CiancioO:04,CiancioPOK:06,Ciancio:06} 
developed unidirectional wavelet transforms on 1D routing paths in WSNs. 
Extensions to 2D routing paths on arbitrary routing trees were made by the authors in~\cite{Shen1,Shen2}.
The superiority of these 1D~\cite{CiancioPOK:06} and 2D~\cite{Shen1} transforms 
over the method in~\cite{Wagner2} (which requires a great deal of backward communication) 
was demonstrated in~\cite{Shen1}. 
%was demonstrated in~\cite{Shen1} in that the amount of energy consumption required to achieve a fixed 
%reconstruction quality of the data is lower than the transform of~\cite{Wagner2}. 
General unidirectional transforms were initially proposed by us in~\cite{Shen4}, in the 
context of lifting transforms~\cite{Sweldens}, and conditions for single-level invertible unidirectional lifting transforms 
were initially proposed there. 
%GSrev2 Re-worded this to point that our previous work did not provide invertibility conditions for general transforms
However, no invertibility conditions were provided for general 
unidirectional transforms, nor were any conditions given for invertible multi-level unidirectional lifting 
transforms. We provide such conditions here (Section~\ref{sec:in-network-processing} and~\ref{sec:uni-separable}) 
as well as new transform designs (Section~\ref{sec:uni-separable-haar}) 
which outperform our previously proposed transforms.

%GSrev1 Moved some of this discussion up to the previous paragraph, right after we state our main contribution
%While existing en-route transforms have been shown to be more efficient, to 
%the best of our knowledge, none of the existing works have attempted to define the most general set of en-route transforms, 
%nor has any attempt been made to find conditions under which such transforms are invertible. 
%The paper is outlined as follows. 
General unidirectional transforms with a set of necessary and sufficient invertibility conditions are 
presented in Section~\ref{sec:in-network-processing}. 
%GSrev1 Moved this statement up to help highlight our contribution
%Our proposed theory also incorporates the use of broadcast data in a general setting. 
%This leads us to develop transforms that use broadcasts in a manner not previously considered. 
In order to demonstrate the generality of our proposed theory, Section~\ref{sec:uni} shows how existing unidirectional 
transforms (e.g., the tree-based KLT~\cite{Shen5}, tree-based differential pulse code modulation (T-DPCM)~\cite{Shen5,senzip} and 
lifting transforms~\cite{Shen4,Shen5}) can be mapped into our framework. 
%lifting transforms proposed in our previous work~\cite{Shen4,Shen5}) can be mapped into our framework. 
Moreover, our proposed formalism is used to construct general unidirectional lifting transforms. 
Some of the inefficiencies of existing lifting transforms are then discussed. 
In order to address these inefficiencies, we define a new 
Haar-like wavelet transform in Section~\ref{sec:uni-separable-haar} which is analogous to the standard Haar wavelet when 
applied to 1D paths. 
As is shown in Section~\ref{sec:uni-separable-haar}, our formalization guarantees invertibility of these Haar-like transforms, 
and also leads to an extension which incorporates broadcast. 
Section~\ref{sec:results} provides experimental results 
that demonstrate the benefits of using our proposed transforms. 

%%%%%%%%%%%%%%%%%%%%%%%%%%%%%%%%%%%%%%%%%%%%%%%%%%%%%%%%%%%%%%%%%%%%%%%%%%%%%%%%%%%%%%%%%%%%%%%%
%%%%%%%%%%%%%%%%%%%%%%%%%%%%%%%%%%%%%%%%%%%%%%%%%%%%%%%%%%%%%%%%%%%%%%%%%%%%%%%%%%%%%%%%%%%%%%%%
%%%%%%%%%%%%%%%%%%%%%%%%%%%%%%%%%%%%%%%%%%%%%%%%%%%%%%%%%%%%%%%%%%%%%%%%%%%%%%%%%%%%%%%%%%%%%%%%
%%%%%%%%%%%%%%%%%%%%%%%%%%%%%%%%%%%%%%%%%%%%%%%%%%%%%%%%%%%%%%%%%%%%%%%%%%%%%%%%%%%%%%%%%%%%%%%%

%\section{The En-route In-network Processing Problem}
\section{En-route In-network Transforms}
\label{sec:in-network-processing}

%This section addresses the main goal of this work, i.e., assuming a fixed routing tree $T$ and transmission schedule $t(n)$ 
%are given, we solve the en-route in-network processing problem. 
In this section, assuming a fixed routing tree $T$ and schedule $t(n)$ are given, 
we provide a definition of unidirectional transforms and determine conditions for their invertibility. 
Some notation is established in Section~\ref{sec:in-network-processing-notations}. 
%We then provide some simplifying assumptions in Section~\ref{sec:in-network-processing-assumptions}. 
Unidirectional transforms are then defined in Section~\ref{sec:in-network-processing-transforms}. 
Section~\ref{sec:in-network-processing-conditions} presents a set of 
conditions under which these transforms are invertible.
%AOrev1 I think we should highlight somewhere that we provide a formalization based on knowledge of the exact time schedule and we later address but then discuss practical issues. May need some editing depending on what is in the end included in that section, as it is it is somewhat ambiguous. 
Throughout this discussion, the configuration of the network in terms routing and scheduling is assumed to be known.  Section~\ref{sec:in-network-processing-discussion} addresses how this can be achieved in practice and how our approach can be used with decentralized initialization approaches. 

\subsection{Notation}
\label{sec:in-network-processing-notations}

Assume there are $N$ nodes in the network with a given routing tree $T = (V, E_T)$, where 
$V = \{1, 2, \ldots, N, N+1\}$, each node is indexed by $n \in \Ic = \{1, 2, \ldots, N\}$, 
the sink node is indexed by $N+1$, and $(m,n) \in E_T$ denotes an edge from 
node $m$ to node $n$ along $T$. We also assume that there is a graph $G = (V,E)$ which is defined by 
the edges in $E_T$ and any additional edges that arise from the broadcast nature of wireless 
communications. An example graph is shown on the right side of Fig.~\ref{fig:BroadcastExample}. 
We observe that data gathering consists of three key components. 
The first is \emph{data measurement}, where each node $n$ measures some scalar data $x(n)$ that it must send to the sink in each epoch 
(these ideas can be easily generalized to non-scalar 
%GSrev5 Added this footnote to describe this generalization in a bit more detail
data\footnote{One straightfoward extension is to use a ``separable'' transform, where a transform is first applied in one dimension (e.g., over time or across dimensions of a multivariate input) and then in the other (i.e., spatially).}). 
%We also assume that data transmissions are scheduled~\cite{Cidon,Sohrabi} to allow sufficient time for data processing, 
%packetization, forwarding, and also to avoid problems like interference and packet collisions. 
Additionally, node $n$ must route its data to the sink along $T$. 
The tree $T$ is defined by assigning to every node $n$ a parent $\rho(n)$. 
We assume that these trees are provided by a standard routing protocol such as CTP. 
Finally, we assume that data transmissions are scheduled~\cite{Cidon,Sohrabi} in some manner, i.e., 
node $n$ will transmit data to its parent $\rho(n)$ at time $t(n)$ according to 
a \emph{transmission schedule} (see Definition~\ref{def:transmission-schedule}). 
CTP is a practical example that can be viewed in terms of this formalization: nodes are assigned parents in a distributed manner, 
data is forwarded to the sink along the corresponding routing paths and 
the times at which nodes transmit serve as an implicit transmission schedule.

\begin{definition}[Transmission Schedule]
\label{def:transmission-schedule}
A transmission schedule 
is a function $t:\Ic \rightarrow \{1, 2, \ldots, M_{slot}\}$, 
such that $t(n) = j$ when node $n$ transmits in the 
$j$-th time slot\footnote{Note that these time slots are not necessarily
 of equal length; they simply allow us 
to describe the order in which communications proceed 
in the network; before
 time slot $t(n)$, node $n$ is listening to other nodes, and at time $t(n)$
 node $n$ starts transmitting its own data, and potentially data from 
its descendants in the routing tree.}. Moreover, node $n$ 
transmits data before node $m$ whenever $t(n) < t(m)$. 
%When the time slots are unique to each node, i.e., $t(n) \neq t(m)$ 
%for all $n \neq m$, we refer to it as a \emph{maximal time slot allocation}. 
\end{definition}

Note that, along the tree $T$, each node has a set of \emph{descendants} $\Dc_n$ which use node $n$ as a data relay to the sink 
and a set of \emph{ancestors} $\Ac_n$ that node $n$ uses for relaying data to the sink. 
%GSrev6 State timing assumption on descendants here
Moreover, we only consider to be descendants of $n$ those nodes that are descendants on the 
tree and transmit earlier than $n$. 
%In order to allow each node to process its own data with data from all of its descendants, 
%we assume that $t(n) > t(m)$ for all $m \in \Dc_n$, i.e., node $n$ transmits after 
%all of its descendants have transmitted. 
Also let each node $n$ be $h(n)$ hops away from the sink node, i.e., $n$ has depth $h(n)$ in $T$. 
We also let $\Cc_n^k$ denote the descendants of $n$ which are exactly $k$ hops away from $n$, 
i.e., $\Cc_n^k = \{m \in \Dc_n | \rho^k(m) = n\}$, where 
$\rho^k(m)$ is the $k$-th ancestor of node $m$ (e.g., $\rho^1(m)$ is the parent of 
$m$, $\rho^2(m)$ is the grandparent of $m$, etc). For instance, $\Cc_n^1$ is the set of 
children of $n$, $\Cc_n^2$ is the set of grandchildren of $n$, etc, and for simplicity we let 
$\Cc_n = \Cc_n^1$. Also note that data can be heard via broadcast in many networks (e.g., WSNs), so 
%GSrev6 Define full set of broadcast neighbors here
we let $\Bc_n^f$ define the \emph{full set of broadcast neighbors} whose data node $n$ can overhear due to broadcast.
%we let $\Bc_n^f$ define the \emph{full set of broadcast neighbors} whose data node $n$ can overhear due to broadcast, 
%i.e., $\Bc_n^f = \{m | (m,n) \in E, m \notin \Cc_n \cup \{\rho(n)\}\}$. 
%In reality, $n$ will receive data pertaining to $\Bc_n$ and also the descendants of each $m \in \Bc_n$, so 
%we also define $\bar{\Bc}_n = \Bc_n \cup_{m \in \Bc_n} \Dc_m$ for the sake of future discussions. 

Under this formulation, each node $n$ can process its own data $x(n)$ together 
with data received from $\Dc_n$ and $\Bc_n^f$. This yields transform coefficient $y(n)$ for node $n$ 
%with data received from descendants and broadcast neighbors. This yields transform coefficient $y(n)$ for node $n$, 
%along with transform coefficients for its descendants $y(\Dc_n)$. Note that node $n$ is only responsible for 
and transform coefficients for its descendants, i.e., $y(m)$ for all $m \in \Dc_n$. 
We make an abuse of notation by letting $y(\Dc_n) = \{y(m) | m \in \Dc_n\}$. Note that node $n$ is only responsible for 
forwarding $y(n)$ and $y(\Dc_n)$ to its parent $\rho(n)$, thus, it should not transmit any data 
received from broadcast neighbors. 
In particular, we assume that node $n$ transmits the \emph{transform coefficient vector} $\yv_n = \left[y(n) \hspace{1mm} y(\Dc_n)\right]^t$ to its parent $\rho(n)$ at time $t(n)$. 
We refer to this as \emph{critical-sampling}, 
%AOrev1 minor rewording along similar lines as the one earlier 
where in each epoch 
only one transform coefficient per sample per node is generated 
and then transmitted to the sink. 
%In other words, each piece of data travels through a unique path to the sink. 
%This is crucial for efficient use of network resources since it prevents nodes from 
%transmitting redundant transform coefficients. 
%Otherwise, the total number of coefficients that are transmitted will increase.
In our formulation, we also allow $y(n)$ (and $y(\Dc_n)$) to 
be further processed at the ancestors of $n$. We refer to this type of processing as \emph{delayed processing}. 

\begin{figure}[htb]
 \centerline{\psfig{figure=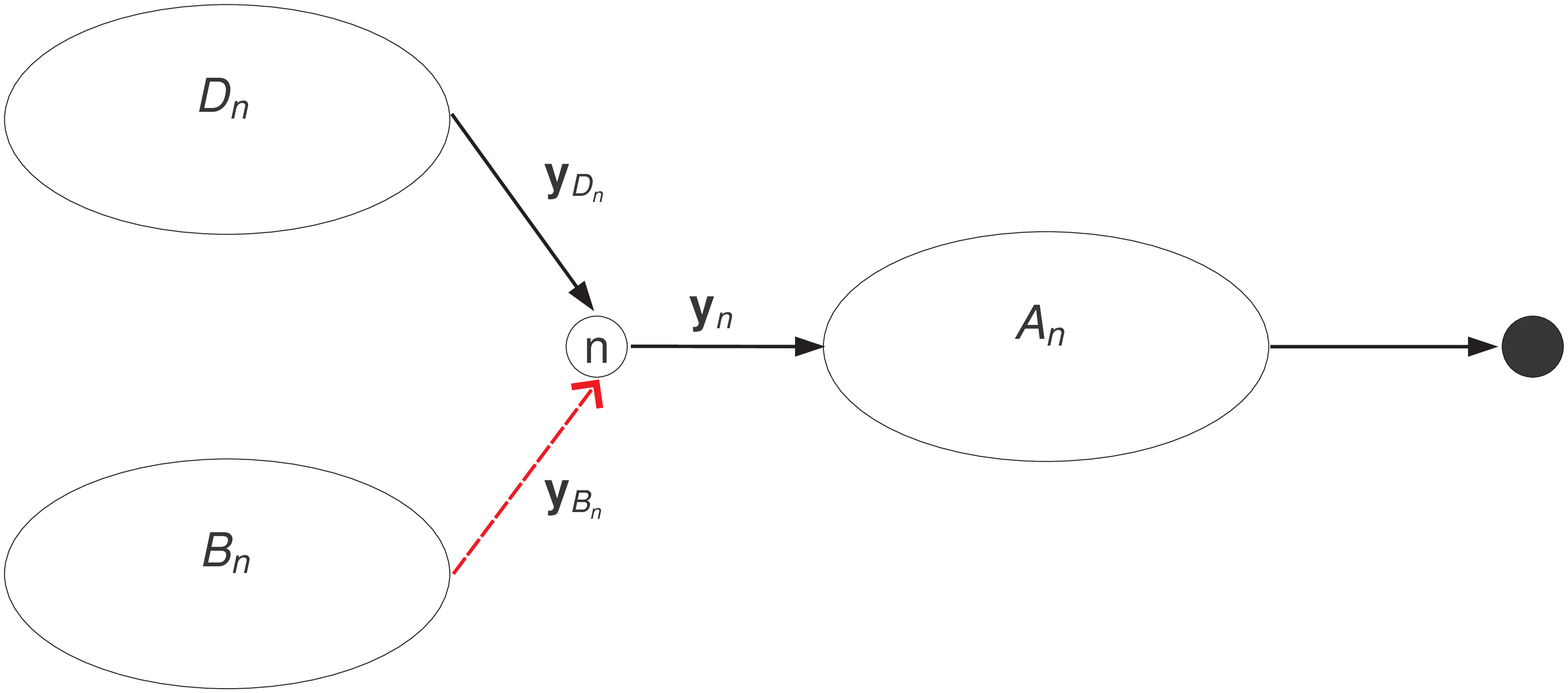,width=8cm,height=4cm}}
  \caption{{\footnotesize Example of causal neighborhoods for each node. 
  Node $n$ receives $\yv_{\Dc_n}$ and $\yv_{\Bc_n}$ from $\Dc_n$ and $\Bc_n$, 
  respectively, processes $x(n)$ together with $\yv_{\Dc_n}$ and $\yv_{\Bc_n}$, 
  then forwards its transform coefficient vector $\yv_n$ through its ancestors in $\Ac_n$. }}
\label{fig:NeighborExample}
\end{figure}

Note that data is only transmitted along $T$ towards the sink, i.e., data relay is \emph{unidirectional} 
towards the sink. 
The existence of a transmission schedule - given explicitly or implicitly - also induces a notion of \emph{causality} 
for transform computations.  In particular, the computations performed at each node $n$ can only involve $x(n)$ and 
any $\yv_{m}$ received from a node $m$ that transmits data before node $n$.
%(i.e., any $\yv_{m}$ such that $t(m) < t(n)$). 
%More specifically, we assume that $t(k) < t(n)$ for all $k \in \Cc_n$ and $t(l) < t(n)$ for all $l \in \Bc_n$, 
%that way every node $n$ can process $x(n)$ together with $\yv_k$ received from each $k \in \Cc_n$ and 
%$\yv_l$ received from each $l \in \Bc_n$. 
%More specifically, we assume that $t(m) < t(n)$ for all $m \in \Dc_n \cup \Bc_n$, 
%GSrev6 Discuss causality here, and how it affects which broadcast neighbors can be used
More specifically, nodes can only use data from $m \in \Bc_n^f$ if $t(m) < t(n)$ (we assume that 
$t(m') < t(n)$ for all $m' \in \Dc_n$). 
%More specifically, we assume that $t(m) < t(n)$ for all $m \in \Cc_n \cup \Bc_n$, 
%that way every node $n$ can process $x(n)$ together with $\yv_m$ received from each $m \in \Cc_n \cup \Bc_n$. 
%that way every node $n$ can process $x(n)$ together with $\yv_m$ received from each $m \in \Cc_n \cup \Bc_n$. 
%This includes data received from direct neighbors of $n$ (i.e., children and broadcast neighbors) as well data that $n$ 
%receives over multi-hop routing paths from descendants. 
%GSrev3 Maybe causal neighborhoods is more appropriate (we used to call them ``tree neighborhoods'')?  I updated the 
%text to reflect this new terminology.
These constraints (i.e., causality and unidirectional relay) induce \emph{causal neighborhoods} whose data 
%GSrev6 Define causal broadcast neighbors here
each node $n$ can use for processing, where we let $\Bc_n = \{m \in \Bc_n^f | t(m) < t(n)\}$ denote 
the \emph{set of causal broadcast neighbors}.
These can be abstracted as in Fig.~\ref{fig:NeighborExample} where 
%for simplicity, we let $\Cc_n = \{c_n^1, c_n^2, \ldots, c_n^{|\Cc_n|}\}$ and 
%$\Bc_n = \{b_n^1, b_n^2, \ldots, b_n^{|\Bc_n|}\}$, and define 
%$\yv_{\Dc_n} = \left[\yv_{c_n^1}^t \hspace{1mm} \hdots \hspace{1mm} \yv_{c_n^{|\Cc_n|}}^t\right]^t$ and 
%$\yv_{\Bc_n} = \left[\yv_{b_n^1}^t \hspace{1mm} \hdots \hspace{1mm} \yv_{b_n^{|\Bc_n|}}^t\right]^t$. 
$\yv_{\Dc_n} = \left[\yv_{\Cc_n(1)}^t \hspace{1mm} \hdots \hspace{1mm} \yv_{\Cc_n(|\Cc_n|)}^t\right]^t$ and 
$\yv_{\Bc_n} = \left[\yv_{\Bc_n(1)}^t \hspace{1mm} \hdots \hspace{1mm} \yv_{\Bc_n(|\Bc_n|)}^t\right]^t$. 
These are formally defined as follows.

%GSrev3 Added a definition of causal neighborhoods here, after introducing timing constraints that are imposed by causality
%AOrev3 This looks good, but here you call them again tree neighborhoods. Also why not come out and say directly that it consists of descendants and broadcast neighbors *such that* transmission happens before ... etc In other words make causality part of the definition of who forms the neighborhood. 
%GSrev6 Redefine causal neighborhoods
\begin{definition}[Causal Neighborhoods]
\label{def:neighbors}
Given a routing tree $T$ and schedule $t(n)$, the \emph{causal neighborhood} of 
each node $n$ is the union of the descendants $\Dc_n$ and the set of causal broadcast neighbors 
$\Bc_n = \{m \in \Bc_n^f | t(m) < t(n)\}$, i.e., $\Dc_n \cup \Bc_n$. 
%Note that $t(m) < t(n)$ for all $m \in \Dc_n \cup \Bc_n$. 
We also define $\bar{\Bc}_n = \Bc_n \cup_{m \in \Bc_n} \Dc_m$ for future discussions. 
\end{definition}

\begin{figure}[htb]
 %\centerline{\psfig{figure=toy_example.eps,width=8cm,height=4cm}}
 %\centerline{\psfig{figure=toy_example.eps,width=12cm,height=6cm}}
 %\centerline{\psfig{figure=toy_example.eps,width=11cm,height=5.5cm}}
 \centerline{\psfig{figure=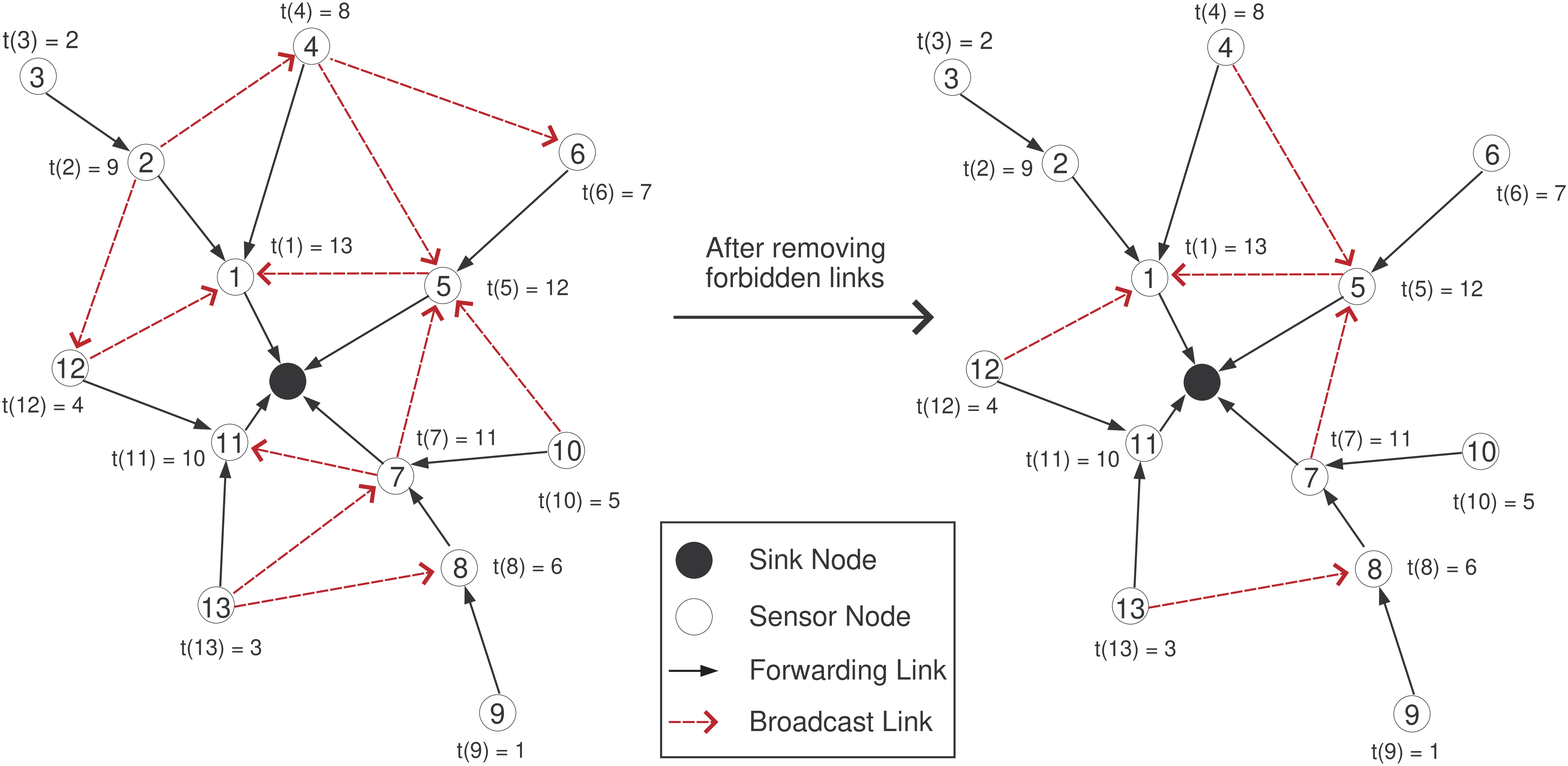,width=10cm,height=5cm}}
  \caption{{\footnotesize Illustration of causal neighborhoods. Node $n$ transmits at time $t(n)$. 
  The left figure shows the full communication graph. The right figure shows the 
  graph after removing broadcast links that violate causality and 
  step by step decoding.}}
 \vspace{-0.3cm}
\label{fig:toy_example}
\end{figure}

These ideas are illustrated in Fig.~\ref{fig:toy_example}. 
%Note how the transmission time of $n$ is always later than that of its descendants $\Dc_n$, i.e., 
%$t(n) \geq t(m)$ for all $m \in \Dc_n$. 
For instance, when node 2 forwards data to node 1, its communication is also overheard 
by nodes 4 and 12. However, nodes 4 and 12 will not receive data from node 2 before they 
transmit, thus, they cannot use it for processing. 

\subsection{Definition of Unidirectional Transforms}
\label{sec:in-network-processing-transforms}

We define a \emph{unidirectional transform} (not necessarily invertible) as 
any transform that (i) is computed unidirectionally along a tree $T$ 
and (ii) satisfies causality and critical sampling. 
%
%We define a \emph{unidirectional transform} (which is not necessarily invertible) as 
%any transform that (i) is computed unidirectionally along a tree $T$ 
%and (ii) satisfies causality and critical sampling. 
Now we can establish the general algebraic form of unidirectional transforms. 
%Now we can establish the general algebraic form of unidirectional transforms 
%(i.e., causal, critically-sampled transforms computed unidirectionally along $T$). 
Without loss of generality, assume that node indices follow a pre-order 
numbering~\cite{Valiente} on $T$, i.e., $\Dc_n = \{n+1, n+2, \ldots, n+|\Dc_n|\}$ for all $n$ 
(see Fig.~\ref{fig:toy_example} for an example of pre-order numbering). 
A pre-order numbering always exists, and can be found via standard algorithms~\cite{Valiente}. 
For the sake of simplicity, we also assume that the transmission schedule $t$ provides 
a unique time slot to each node\footnote{We note that the time slot assignment need not be unique. However, 
this assumption significantly simplifies the transform construction and invertibility conditions. 
It is easy to develop similar transform constructions when multiple nodes are assigned the same 
time slots, and similar invertibility conditions arise.}, 
i.e., $t(n) \neq t(m)$ for all $n \neq m$. 
%i.e., $M_{slot} = N$ and $t(n) \neq t(m)$ for all $n \neq m$. 

%Let $\xv = \left[x(1) \hspace{1mm} x(2) \hspace{1mm} \hdots \hspace{1mm} x(N)\right]^t$ 
%be the vector of raw data and let $\yv_n$ be the vector of transform coefficients 
%generated (and transmitted) by node $n$. Refer again to Fig.~\ref{fig:NeighborExample} and note 
Recall that each node $n$ receives $\yv_{\Dc_n}$ and $\yv_{\Bc_n}$ from its descendants 
and (causal) broadcast neighbors, respectively (see Fig.~\ref{fig:NeighborExample}).
%Let the data received from descendants be expressed as an aggregate vector $\yv_{\Dc_n}$. 
%Also define $\yv_{\Bc_n}$ as the vector which contains data received from broadcast neighbors and their descendants, 
%i.e., data from $\bar{\Bc}_n$.  For $\Bc_n = \{b_n^1, b_n^2, \ldots, b_n^{|\Bc_n|}\}$, we have that 
%%$\yv_{\Bc_n} = (\yv_{b_n^1}^t \hspace{1mm} \yv_{b_n^2}^t \hspace{1mm} \hdots \hspace{1mm} \yv_{b_n^{|\Bc_n|}}^t)^t$. 
%$\yv_{\Bc_n} = \left[ \yv_{b_n^1}^t \hspace{1mm} \yv_{b_n^2}^t \hspace{1mm} \hdots \hspace{1mm} \yv_{b_n^{|\Bc_n|}}^t \right]^t$. 
Thus, in a general unidirectional transform, each node $n$ processes its own 
data $x(n)$ along with $\yv_{\Dc_n}$ and $\yv_{\Bc_n}$. 
Then, it will transmit transform coefficient vector $\yv_n$ at 
time $t(n)$. We omit $t(n)$ from the notation of 
$\yv_n$ since the timing is implicit. 
In order to satisfy critical-sampling, it is necessary 
that each node only forward $1+|\Dc_n|$ coefficients to the sink. 
Therefore, $\yv_n$ must be a $(1+|\Dc_n|) \times 1$ dimensional vector. 
A unidirectional transform can now be expressed as follows.
%Under this formulation, a unidirectional transform can be 
%expressed as follows.

%GSrev2 Reworded the definition to reflect the more general formulation without timing assumptions
%AOrev2 I'm confused. Yes, there are no timing assumptions made but some are implicitly made, since we use data from descendants and broadcast neighbors. If we define the transform this way, doesn't that imply that those neighbors transmit first? Is it clear from the previous section that descendants and broadcast neighbors are only those that transmit before node n? It might be worth emphasizing this (even in the definition itself). 
%GSrev3 Now make reference to new definition of causal neighborhoods
\begin{definition}[Unidirectional Transform]
\label{def:transform}
Let $T$ be a routing tree with a unique time slot assignment given by $t(n)$, 
and suppose that the causal neighborhood of each node is given by 
Definition~\ref{def:neighbors}. 
%and suppose that the assumptions in 
%Section~\ref{sec:in-network-processing-notations} are satisfied. 
A unidirectional transform on $T$ is a collection of local transformations 
done at each node $n$ given by
\begin{equation}
  \yv_n = \left[ \Am_n \hspace{1mm} \Bm_n^1 \hspace{1mm} \hdots \hspace{1mm} \Bm_n^{|\Bc_n|} \right] \cdot 
          \left[ \begin{array}{c} x(n) \\ \yv_{\Dc_n} \\ \yv_{\Bc_n} \end{array} \right], 
  \label{eqn:transform}
\end{equation}
where $\yv_n$ has dimension $(1+|\Dc_n|) \times 1$, $\Am_n$ has 
dimension $(1+|\Dc_n|) \times (1+|\Dc_n|)$ and 
each $\Bm_n^i$ has dimension $(1+|\Dc_n|) \times (1+|\Dc_{\Bc_n(i)}|)$. 
The transform is computed starting from the node at the first time slot 
up through the nodes in the remaining time slots $k = 2, 3, \ldots, N$. 
\end{definition}

\subsection{Invertibility Conditions for Unidirectional Transforms}
\label{sec:in-network-processing-conditions}

%GSrev2 Moved the discussion on step by step decodability and timing assumptions here
We now establish a set of invertibility conditions for unidirectional transforms. 
Note that these transforms are always computed in a particular order, e.g., 
starting from nodes furthest from the sink (i.e., ``leaf'' nodes), up to 
nodes which are 1-hop from the sink. Some sort of interleaved scheduling (where 
one set of nodes transmits before the rest) could 
also be used~\cite{sunil2}. Therefore, it would also be desirable to have step by step 
decoding in the reverse order, 
since this would simplify the transform constructions. 
In particular, if the overall transform can be inverted by inverting the computations done at each node in 
the reverse order, then invertibility will be ensured by designing invertible transforms at each node. 

Step by step decoding in the reverse order is 
trivially guaranteed when no broadcast data is used since the transform 
at each node $n$ is simply 
$\yv_n = \Am_n \cdot \left[x(n) \hspace{1mm} \yv_{\Dc_n}^t\right]^t$. 
Thus, if each $\Am_n$ is invertible, we can invert the operations 
done at node $n$ as 
$\left[x(n) \hspace{1mm} \yv_{\Dc_n}^t\right]^t = \left(\Am_n\right)^{-1} \cdot \yv_n$. 
This becomes more complicated when broadcast data is used. 
By examining (\ref{eqn:transform}), we observe that 
$\yv_n = \Am_n \cdot \left[x(n) \hspace{1mm} \yv_{\Dc_n}^t \right]^t + \left[ \Bm_n^1 \hspace{1mm} \hdots \hspace{1mm} \Bm_n^{|\Bc_n|} \right] \cdot \yv_{\Bc_n}$, 
where $\yv_{\Bc_n} = \left[\yv_{\Bc_n(1)}^t \hspace{1mm} \hdots \hspace{1mm} \yv_{\Bc_n(|\Bc_n|)}^t\right]^t$. 
In order to have step by step decodability, we need to be able to recover (for every node $n$) 
$x(n)$ and $\yv_{\Dc_n}$ from $\yv_n$ and $\yv_{\Bc_n}$. 
Note that this fails whenever we 
cannot decode some transform coefficient vector $\yv_{m}$ from broadcast node $m \in \Bc_n$ 
%cannot decode some transform coefficient vector $\yv_{\Bc_n(i)}$ from node $\Bc_n(i)$ 
before decoding $\yv_n$. It will also fail if the 
matrix operations performed at any given node are not invertible. 
Thus, in order to guarantee step by step decodability, we need to ensure that 
(i) the matrix operations at each node are invertible, and 
(ii) it is possible to decode each $\yv_{m}$ before decoding $\yv_n$. 
%(ii) it is possible to decode each $\yv_{\Bc_n(i)}$ before decoding $\yv_n$. 
As we now show, (i) is guaranteed by ensuring that 
each $\Am_n$ matrix is invertible and 
(ii) is guaranteed by imposing a timing condition.

\begin{proposition}[Step by Step Decodability]
Suppose that we have the transform in Definition~\ref{def:transform} and assume that 
$t(\rho(m)) > t(n)$ for every broadcast node $m \in \Bc_n$. 
Then we can recover $x(n)$ and $\yv_{\Dc_n}$ as 
%$\left[x(n) \hspace{1mm} \yv_{\Dc_n}^t \right]^t = 
%\left(\Am_n\right)^{-1} \cdot \yv_n - \left(\Am_n\right)^{-1} \cdot \left[ \Bm_n^1 \hspace{1mm} \hdots \hspace{1mm} \Bm_n^{|\Bc_n|} \right] \cdot \yv_{\Bc_n}$, i.e., step by step coding is guaranteed.
$\left[x(n) \hspace{1mm} \yv_{\Dc_n}^t \right]^t = 
\left(\Am_n\right)^{-1} \cdot \yv_n - \left(\Am_n\right)^{-1} \cdot \left[ \Bm_n^1 \hspace{1mm} \hdots \hspace{1mm} \Bm_n^{|\Bc_n|} \right] \cdot \yv_{\Bc_n}$ if and only if $\Am_n^{-1}$ exists.
%Moreover, (i) is a necessary condition for step by step decodability.
\label{prop:conditions1}
\end{proposition}

\begin{IEEEproof}
Note that the vector transmitted by any broadcast node $m \in \Bc_n$ will be processed at its parent, 
node $\rho(m)$, and this processing will occur at time $t(\rho(m))$. 
Moreover, node $n$ will generate its own transform coefficient vector $\yv_n$ at time $t(n)$, 
and by assumption we have that $t(\rho(m)) > t(n)$. Thus, it is possible to decode 
$\yv_m$ before $\yv_n$ for every broadcast neighbor $m \in \Bc_n$. Thus, we can always form 
$\yv_{\Bc_n} = \left[\yv_{\Bc_n(1)}^t \hspace{1mm} \hdots \hspace{1mm} \yv_{\Bc_n(|\Bc_n|)}^t\right]^t$ 
before decoding $\yv_n$. Therefore, 
%a simple manipulation of (\ref{eqn:transform}) will give
we can recover $x(n)$ and $\yv_{\Dc_n}$ as 
$\left[x(n) \hspace{1mm} \yv_{\Dc_n}^t \right]^t = 
\left(\Am_n\right)^{-1} \cdot \yv_n - \left(\Am_n\right)^{-1} \cdot \left[ \Bm_n^1 \hspace{1mm} \hdots \hspace{1mm} \Bm_n^{|\Bc_n|} \right] \cdot \yv_{\Bc_n}$ 
if and only if $\Am_n^{-1}$ exists.
%Clearly, if $\Am_n$ is not invertible, then we cannot recover $x(n)$ and $\yv_{\Dc_n}$. 
%Thus, (i) is also a necessary condition.
\end{IEEEproof}

To simplify our transform constructions, we \emph{also assume} that nodes use the latest version of 
broadcast data that they receive, i.e., $m \in \Bc_n$ only if $\Ac_m \cap \Bc_n = \emptyset$. 
This \emph{second constraint} precludes the possibility that a node $n$ receives broadcast 
data from node $m$ and from an ancestor of node $m$.
Removing the broadcast links which violate these constraints gives a simplified communication 
graph as shown on the right side of Fig.~\ref{fig:toy_example}. 
%AOrev2 Above statement sounds straightforward in theory, but we should make sure that we can explain that this is also feasible in practice and does not require a huge amount of coordination. 
%GSrev3 Added text here per our discussion
%AOrev3 minor edit 
Removal of these links can be done by local information exchange 
within the network; examples of how this can be achieved are 
discussed in Section~\ref{sec:in-network-processing-discussion}.
Under the constraint of Prop.~\ref{prop:conditions1} and this second constraint, 
we can represent the global transform taking place in the network as follows. 
%The local transform performed at node $n$ is given by the matrix 
%$\left[ \Am_n \hspace{1mm} \Bm_n^1 \hspace{1mm} \hdots \hspace{1mm} \Bm_n^{|\Bc_n|} \right]$. 
Since the time slot assignment is unique, at time $t(n)$ only data from $n$ and its descendants 
will be modified, i.e., only $x(n)$ and $y(\Dc_n)$ will be changed at time $t(n)$. 
Since pre-order indexing is used, we have that 
%$\yv_{\Dc_n} = [y(n+1) \hspace{1mm} y(n+2) \hspace{1mm} \hdots \hspace{1mm} y(n+|\Dc_n|)]$. 
$\yv_{\Dc_n} = [y(n+1), \hspace{1mm} \hdots, \hspace{1mm} y(n+|\Dc_n|)]^t$. 
Therefore, the global transform computations done at time $t(n)$ are given by 
(\ref{eq:global-k}), where each $\tilde{\yv}_i$ corresponds to data which is not processed at time $t(n)$. 
%AOrev1 The equation below is a bit complicated by the notation of the vectors (transformed and output). Primarily what's a bit difficult to follow is the vector notation, so for example y_{k+1} seems to be vector, contain also the descendants of k+1, but then with that notation, shouldn't that also include some descendants of y_n (potentially). On the other hand the matrix is very clear (in Eq. (3)). One possible simplification would be to keep only the elements that change in the input/output vectors of (2), and replace everything else by ... then we can say that this matrix operates on all the y values, but only those shown participate in the computation.  
\begin{equation}
\left[ \begin{array}{c}
  \tilde{\yv}_1 \\ \yv_{\Bc_n(1)} \\ \vdots \\ \tilde{\yv}_k \\ \yv_n \\ \tilde{\yv}_{k+1} \\ \vdots \\ \yv_{\Bc_n(|\Bc_n|)} \\ \tilde{\yv}_K
  \end{array} \right]
  = \left[ \begin{array}{ccccccccc}
                 \Id & {\bf 0} & \hdots & {\bf 0} & {\bf 0} & {\bf 0} & \hdots & {\bf 0} & {\bf 0} \\
                 {\bf 0} & \Id & \hdots & {\bf 0} & {\bf 0} & {\bf 0} & \hdots & {\bf 0} & {\bf 0} \\
                 \vdots & \vdots & \ddots & \vdots & \vdots & \vdots & \ddots & \vdots & \vdots \\
                 {\bf 0} & {\bf 0} & \hdots & \Id & {\bf 0} & {\bf 0} & \hdots & {\bf 0} & {\bf 0} \\
                 {\bf 0} & \Bm_n^1 & \hdots & {\bf 0} & \Am_n & {\bf 0} & \hdots & \Bm_n^{|\Bc_n|} & {\bf 0} \\
                 {\bf 0} & {\bf 0} & \hdots & {\bf 0} & {\bf 0} & \Id & \hdots & {\bf 0} & {\bf 0} \\
                 \vdots & \vdots & \ddots & \vdots & \vdots & \vdots & \ddots & \vdots & \vdots \\
                 {\bf 0} & {\bf 0} & \hdots & {\bf 0} & {\bf 0} & {\bf 0} & \hdots & \Id & {\bf 0} \\
                 {\bf 0} & {\bf 0} & \hdots & {\bf 0} & {\bf 0} & {\bf 0} & \hdots & {\bf 0} & \Id \\
  \end{array} \right] 
  %\cdot 
  \left[ \begin{array}{c}
  \tilde{\yv}_1 \\ \yv_{\Bc_n(1)} \\ \vdots \\ \tilde{\yv}_k \\ 
  \left[ \begin{array}{c} x(n) \\ \yv_{\Dc_n} \end{array} \right] \\ 
  %\left( \begin{array}{c} x(n) \\ \yv_{\Dc_n} \end{array} \right) \\ 
  \tilde{\yv}_{k+1} \\ \vdots \\ \yv_{\Bc_n(|\Bc_n|)} \\ \tilde{\yv}_K
  \end{array} \right]
 \label{eq:global-k}
\end{equation}
The \emph{global transform matrix} $\Cm_{t(n)}$ at time 
$t(n)$ is just the matrix shown in (\ref{eq:global-k}), i.e., 
%The \emph{global transform matrix} $\Cm_{t(n)}$ at time 
%$t(n)$ can then be expressed in terms of the local transform by placing 
%$\Am_n$ and each $\Bm_n^j$ in the corresponding locations of $\Cm_{t(n)}$, i.e., 
\begin{equation}
  \Cm_{t(n)} = \left[ \begin{array}{ccccccccc}
                 \Id & {\bf 0} & \hdots & {\bf 0} & {\bf 0} & {\bf 0} & \hdots & {\bf 0} & {\bf 0} \\
                 {\bf 0} & \Id & \hdots & {\bf 0} & {\bf 0} & {\bf 0} & \hdots & {\bf 0} & {\bf 0} \\
                 \vdots & \vdots & \ddots & \vdots & \vdots & \vdots & \ddots & \vdots & \vdots \\
                 {\bf 0} & {\bf 0} & \hdots & \Id & {\bf 0} & {\bf 0} & \hdots & {\bf 0} & {\bf 0} \\
                 {\bf 0} & \Bm_n^1 & \hdots & {\bf 0} & \Am_n & {\bf 0} & \hdots & \Bm_n^{|\Bc_n|} & {\bf 0} \\
                 {\bf 0} & {\bf 0} & \hdots & {\bf 0} & {\bf 0} & \Id & \hdots & {\bf 0} & {\bf 0} \\
                 \vdots & \vdots & \ddots & \vdots & \vdots & \vdots & \ddots & \vdots & \vdots \\
                 {\bf 0} & {\bf 0} & \hdots & {\bf 0} & {\bf 0} & {\bf 0} & \hdots & \Id & {\bf 0} \\
                 {\bf 0} & {\bf 0} & \hdots & {\bf 0} & {\bf 0} & {\bf 0} & \hdots & {\bf 0} & \Id \\
  \end{array} \right].
  \label{eqn:global-transform-k}
\end{equation}
This yields the \emph{global transform coefficient vector}
\begin{equation}
  \yv = \Cm_N \cdot \Cm_{N-1} \cdots \Cm_1 \cdot \xv.
  \label{eqn:global-transform}
\end{equation}

\begin{figure}[htb]
 \centerline{\psfig{figure=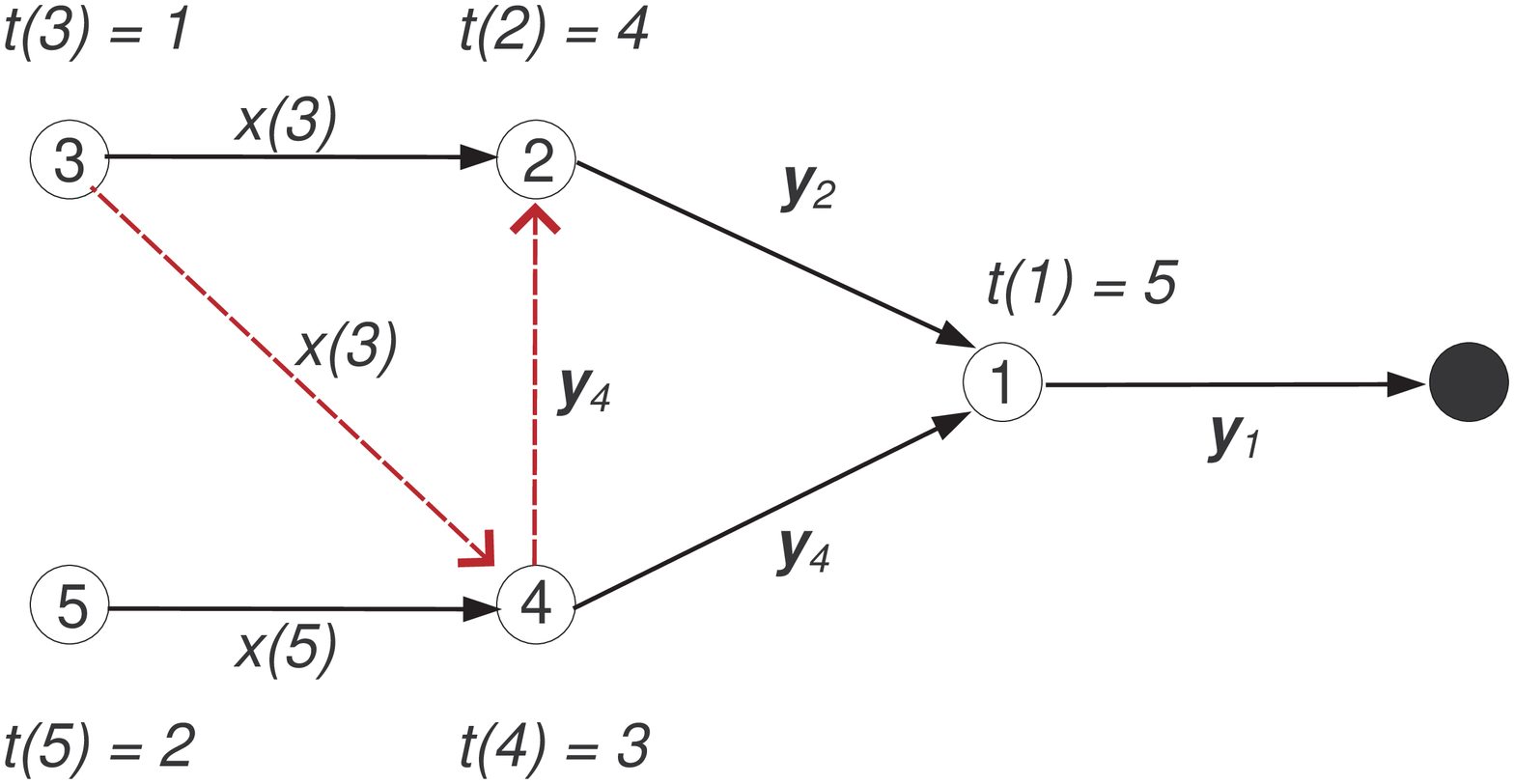,width=8cm,height=4cm}}
   \caption{{\footnotesize Example to illustrate unidirectional transform computations. 
   Nodes generate and transmit transform coefficients in the order specified by the transmission schedule.}}
\label{fig:ConditionsExample}
\end{figure}

Fig.~\ref{fig:ConditionsExample} illustrates these transform computations. 
Initially, $\yv = \xv = \left[x(1) \hspace{1mm} x(2) \hspace{1mm} \hdots \hspace{1mm} x(5)\right]^t$. 
At times 1 and 2, nodes 3 and 5, respectively, transmit raw data to their parents. Therefore, the 
global matrices at times 1 and 2 are simply $\Cm_1 = \Cm_2 = \Id$. At time 3, node 4 
produces
%produces transform coefficients $y(4)$ and $y(5)$ 
%(and coefficient vector $\yv_4$) as 
\begin{equation*}
\yv_4 = \left[\begin{array}{c} y(4) \\ y(5) \end{array}\right] = 
		\left[\begin{array}{ccc} b_1 & a_1 & a_2 \\
		                         b_2 & a_3 & a_4 
		\end{array}\right] 
		\cdot
		\left[\begin{array}{c} x(3) \\ x(4) \\ x(5) \end{array}\right],
%\yv_4 = \left[\begin{array}{c} y(4) \\ y(5) \end{array}\right] = 
%		\left[\begin{array}{ccc} a_1 & a_2 & b_1 \\
%		                         a_3 & a_4 & b_2 
%		\end{array}\right] 
%		\cdot
%		\left[\begin{array}{c} x(4) \\ x(5) \\ x(3) \end{array}\right].
\end{equation*}
where $a_i$ and $b_i$ represent arbitrary values of the 
transform matrix used at node 4.
Then at time 4, node 2 produces transform coefficients $y(2)$ and $y(3)$ 
(and coefficient vector $\yv_2$) as 
\begin{equation*}
\yv_2 = \left[\begin{array}{c} y(2) \\ y(3) \end{array}\right] = 
		\left[\begin{array}{cccc} a_1' & a_2' & b_1' & b_2' \\
		                         a_3' & a_4' & b_3' & b_4' 
		\end{array}\right] 
		\cdot
		\left[\begin{array}{c} x(2) \\ x(3) \\ y(4) \\ y(5) \end{array}\right],
\end{equation*}
where $a_i'$ and $b_i'$ are the values of the matrix used at node 2.
Node 1 then computes $\yv_1$ at time 5. The global transform is 
given by
\begin{equation}
  \yv = \Am_1 
        \left[ \begin{array}{ccccc} 1 & 0 & 0 & 0 & 0 \\ 
                    0 & a_1' & a_2' & b_1' & b_2' \\
                    0 & a_3' & a_4' & b_3' & b_4' \\
                    0 & 0 & 0 & 1 & 0 \\ 
                    0 & 0 & 0 & 0 & 1 \\ 
                    \end{array} \right] 
        \left[ \begin{array}{ccccc} 1 & 0 & 0 & 0 & 0 \\ 
                    0 & 1 & 0 & 0 & 0 \\ 
                    0 & 0 & 1 & 0 & 0 \\
                    0 & 0 & b_1 & a_1 & a_2 \\
                    0 & 0 & b_2 & a_3 & a_4 \\
                    \end{array} \right] 
        \left[ \begin{array}{c} x(1) \\ x(2) \\ x(3) \\ x(4) \\ x(5) \end{array} \right].
%  \yv = \Am_1 \left[ \begin{array}{ccccc} 1 & 0 & 0 & 0 & 0 \\ 
%                    0 & \Am_2(1,1) & \Am_2(1,2) & \Bm_2^1(1,2) & \Bm_2^1(1,2) \\
%                    0 & \Am_2(2,1) & \Am_2(2,2) & \Bm_2^1(2,2) & \Bm_2^1(2,2) \\
%                    0 & 0 & 0 & 1 & 0 \\ 
%                    0 & 0 & 0 & 0 & 1 \\ 
%                    \end{array} \right] 
%        \left[ \begin{array}{ccccc} 1 & 0 & 0 & 0 & 0 \\ 
%                    0 & 1 & 0 & 0 & 0 \\ 
%                    0 & 0 & 1 & 0 & 0 \\
%                    0 & 0 & \Bm_4^1(1,1) & \Am_4(1,1) & \Am_4(1,2) \\
%                    0 & 0 & \Bm_4^1(2,1) & \Am_4(2,1) & \Am_4(2,2) \\
%                    \end{array} \right] 
%        \xv.
  \label{eqn:transform-example}
\end{equation}

It is now rather simple to show that the transform is invertible as long as each matrix $\Am_n$ is invertible. 
%This is formally proven in the following proposition. 
%Given the unidirectional transform construction in Def.~\ref{def:transform} along with the 
%global transform constructions in (\ref{eqn:global-transform-k}) and (\ref{eqn:global-transform}), 
%the transform is invertible as long as each matrix $\Am_n$ is invertible. This is formally proven as follows. 
%the transform is invertible as long as the rows in the matrix 
%$\left[ \Am_n \hspace{1mm} \Bm_n^1 \hspace{1mm} \hdots \hspace{1mm} \Bm_n^{|\Bc_n|} \right]$ 
%are linearly independent (L.I.) for each node $n$. This is formally proven as follows.

%GSrev2 Redefined this a bit to reflect the new structure of this section
%GSrev3 I modified the statement of the proposition and the proof a bit
\begin{proposition}[Invertible Unidirectional Transforms]
\label{prop:invertibility}
%Let $T$ define a set of multi-hop routing paths along which nodes forward 
%data to a single sink node. Let the transmission schedule $t$ define the 
%data transmission times and suppose that it gives a unique time slot to each node. 
Suppose that we have the transform in Def.~\ref{def:transform}, 
%suppose the two broadcast constraints are met 
%%GSrev3 Explicitly state the broadcast conditions we use here
%(i.e., $t(n) < t(m)$ for all $m \in \Bc_n$ and $m \in \Bc_n$ only if $\Ac_m \cap \Bc_n = \emptyset$) 
the second timing constraint ($m \in \Bc_n$ only if $\Ac_m \cap \Bc_n = \emptyset$) is met, 
and Prop.~\ref{prop:conditions1} is satisfied for every node $n$. 
%and let the corresponding global 
%transforms be given by (\ref{eqn:global-transform-k}) and (\ref{eqn:global-transform}). 
%Then the overall transform is invertible if and only if 
%%$\Am_n$ is invertible for every $n$.
%$\Am_n$ is invertible for every node $n$.
%$\left[ \Am_n \hspace{1mm} \Bm_n^1 \hspace{1mm} \hdots \hspace{1mm} \Bm_n^{|\Bc_n|} \right]$ 
%has linearly independent rows for every node $n$.
Then the overall transform given by (\ref{eqn:global-transform}) is invertible.
\end{proposition}

\begin{IEEEproof}
%GSrev3 Added a line here which explains that (\ref{eqn:global-transform}) arises from the two broadcast timing assumptions
Under the two broadcast timing assumptions, the global transform is given by (\ref{eqn:global-transform}). 
(\ref{eqn:global-transform}) is invertible if and only if 
every $\Cm_{t(n)}$ in (\ref{eqn:global-transform-k}) is invertible. 
$\Cm_{t(n)}$ is invertible if and only if $\det\left(\Cm_{t(n)}\right) \neq 0$. 
Recall that adding a multiple of one row to another does not change 
the determinant~\cite{Strang}. 
%AOrev1 added this to emphasize that this is true because of the structure of these matrices 
Given the structure of the $\Cm_{t(n)}$ matrices, 
using such row operations to eliminate each 
$\Bm_n^i$ matrix, it is easy to show that $\det\left(\Cm_{t(n)}\right) = \det\left(\Am_n\right)$. 
Moreover, Prop.~\ref{prop:conditions1} implies that $\Am_n$ is invertible.
\end{IEEEproof}

Proposition~\ref{prop:invertibility} shows that locally invertible 
transforms provide globally invertible transforms. 
Moreover, under our stated timing constraints, broadcast data does not affect invertibility. 
Therefore, broadcast data at each node $n$ can be used in an arbitrary manner without affecting invertibility. 
So in order to design an invertible unidirectional transform, all that one must do 
is design invertible matrices $\Am_n$. 
This is an encouraging result since it essentially means that broadcast data 
can be used in any way a node chooses. In particular, broadcast data can always 
be used to achieve more data decorrelation. 

%GSrev1 Added a new section here to discuss practical issues
\subsection{Discussion}
\label{sec:in-network-processing-discussion}

%AOrev1 This section would work better if we started by explaining what issues were not taken into consideration in the previous sections (i.e., formulating the problem we are solving here). The structure of the section also makes it have a piecemeal feel to it. It is not clear for example why channel effects and packet losses should be different (the latter could be due to congestion). Also, it is not clear if all these practical issues arise simultaneously, if they can be all solved together, etc. It may be easier if we could describe some general problems (initialization, reconfiguration, robustness to losses) and then explain how these may affect our formulation. 

%GSrev2 I attempted a re-write this whole section in accordance with your comments

The theory presented thus far assumes that the routing 
and transmission scheduling are known, and that all of the 
transform matrices are known both at the nodes and at the sink. 
In practice, the routing, scheduling and transforms must be 
initialized. 
Moreover, the network may need to re-configure itself if, for example, 
nodes die or link conditions change drastically. 
In addition, packet losses will often occur. Nodes typically deal with 
this (as in CTP) by re-transmitting a packet until an acknowledgement (ACK) is 
received from the intended recipient. 
While these three issues pose no significant problems 
for routing, they all have an impact on our proposed 
transform due to the assumptions we make about timing. We now provide 
some discussion of how this affects our theory and how it can be handled.
%some discussion of how this affects our theory and propose 
%methods that allow us to work around these issues.

We first address the impact that initialization and reconfiguration have 
on the routing and scheduling, as well as what can be 
done to address it. 
We assume that routing is initialized and reconfigured in a distributed manner 
using standard protocols such as CTP. 
Distributed scheduling protocols for WSNs also exist~\cite{Sohrabi,Avinash}. 
However, the resulting schedules may not be consistent 
with Definition~\ref{def:neighbors} (i.e., 
they may not provide timings for which $t(m) < t(n)$ for all $m \in \Dc_n$), 
so in practice we would need to enforce such timings. One way to 
achieve this is to force nodes to suppress 
transmission (in a given epoch) until they have received data from all of their 
descendants. Another alternative would be to determine such a transmission schedule 
at the sink, then to disseminate the timing information to the nodes. 

Whenever timing and routing information is established (or
re-established due to re-configuration), it is also necessary to check
our main broadcast timing constraint, i.e., $m \in \Bc_n$ only if
$t(n) < t(\rho(m))$.  We describe one way in which this information
can be disseminated to each node in a distributed manner.
%GSrev3 Actually, this can be done using only two messages, as follows
%GSrev3 Old text shown below
%First, each broadcasting node $m$ sends a request message to 
%its parent $\rho(m)$ for its transmission time $t(\rho(m))$, then its parent sends a message containing 
%$t(\rho(m))$ back to node $m$. Finally, $m$ broadcasts a message to its neighbors containing $t(\rho(m))$. 
%GSrev3 New text shown below
First, whenever the time $t(n)$ at node $n$ is initialized or changes,
it broadcasts a small packet (i.e., a \emph{beacon}) which contains
$t(n)$ to its children. Then, any child of $n$ which broadcasts data
will send the same beacon to all of its neighbors.
%AOrev2 The concern here is that in a sensor network setting, due to packet headers and other overheads, the difference between small and large messages is relatively small. Can some of this be done piggybacking on the original set of packet transmissions?
%GSrev3 Reworded this to describe how we can piggyback on what protocols like CTP do already
This requires a total of 2 messages for each broadcasting node. 
%This will incur an additional cost, though it should be relatively minor compared with the cost for 
%data transmission. There may be better alternatives to this, and one could always choose not to use broadcast 
%data in a real network to avoid this overhead, but such decisions are beyond the scope of this work. 
%GSrev3 New text shown below
Note that protocols such as CTP already use control beacons (in addition to data packets) 
to update stale routing information. 
Thus, nodes could potentially piggyback timing information on these control beacons whenever they are generated, 
or otherwise use separate control beacons to disseminate timing information. 
%Timing information can also be disseminated using separate control beacons whenever nodes cannot piggyback it. 
This will incur an additional cost, although (as was shown in~\cite{ctp}) the per packet cost for control beacons 
is typically much smaller than the cost for data forwarding.

Initialization and re-configuration also impacts the 
transform matrices that are used. 
Each node could transmit the values of its matrix to the sink, 
or viceversa, but this may be very costly. Instead, 
the construction of each transform matrix should be 
based on a small amount of information which is made common 
to the nodes and to the sink. For example, 
the values in each transform matrix could be based on the number of 1-hop neighbors that 
each node has~\cite{Shen1} or the relative node positions~\cite{Wagner1}. 
In this way each matrix can be constructed at each node and at the sink 
without explicitly communicating the matrix values. However, 
additional information (e.g., node positions, number of neighbors) 
would need to be communicated to the sink whenever the network is initialized or re-configured.
%AOrev2 we could give a very simple example: e.g., use only the number of nodes used, and send as overhead which nodes were used. 
%GSrev3 Added a simple example here
For example, each node could construct a transform using only the number of nodes that it 
receives data from (as in~\cite{Shen1,Wagner1}) and would send the set of nodes whose data it used as overhead to the sink. 
Then, assuming that the nodes and the sink construct the matrices according to the same rules, the 
sink can re-construct the matrix used at each node.

Packet loss is the last practical issue which impacts our proposed transforms. 
%AOrev2 saying that it's out of scope is a little too weak. a more accurate characterization is that there are lots of mechanisms that can handle channel noise, and we take advantage of these. 
%GSrev3 Okay, reworded accordingly
We do not consider the effects of channel noise on the data since these can be handled using a wide variety of existing techniques. 
Moreover, packet losses and channel noise will impact other data gathering schemes (e.g., CTP), 
and we expect that the penalty due to packet losses will be similar in our scheme and in other data gathering schemes. 
Packet losses are typically handled (as in CTP) by re-transmitting a packet until an ACK is received from the desired destination. 
Thus, if node $n$ does not receive data from descendant $n+k$ by the time that it transmits, due to packet re-transmissions for $n+k$, 
the data from node $n+k$ cannot be combined with data available at node $n$. 
This is equivalent to not using the data from node $n+k$ in the transform computation 
(i.e., $\Am_n(j,k+1) = \Am_n(k+1,j) = 0$ for all $j \neq k+1$ and $\Am_n(k+1,k+1) = 1$) and 
does not affect our proposed theory. 
However, this change must be signaled to the sink so that it knows how to 
adjust $\Am_n$ accordingly. This can be done by including some additional information in the packet headers 
for node $n$ and $n+k$ to signify this change. 

Packet losses also have an impact on the use of broadcast data. Suppose that node $n$ does not 
receive a data packet from broadcast neighbor $b_k$ but the packet from $b_k$ does reach the intended recipient $\rho(b_k)$. 
In this case, node $\rho(b_k)$ will send an ACK back to node $b_k$ and node $b_k$ will no longer 
re-transmit (note that node $b_k$ will not expect an ACK from node $n$). 
Thus, data from node $b_k$ can not be combined with data available at node $n$. 
This is equivalent to not using data from node $b_k$ in the transform computation (i.e., $\Bm_n^k = \bf{0}$) and our proposed 
theory is not affected. However, this change must be signaled to the sink so that it knows to set $\Bm_n^k = \bf{0}$. 

One way to work around these issues (initialization, re-configuration and packet losses) 
is to design transforms that can work under arbitrary timing and with arbitrary 
uses of broadcast data. 
However, under arbitrary timing and use of broadcast data, 
it is no longer possible to guarantee 
global transform invertibility by designing invertible transforms at each node. More specifically, 
we must ensure that the transform computations done at different nodes are jointly invertible. 
This leads to a set of complex conditions. The cost to determine such conditions and to 
coordinate nodes so that they satisfy these conditions could be very high, perhaps even much 
higher than the additional coordination needed to implement our proposed transforms. 
However, it is still possible to design simple versions of such transforms by using 
constructions such as lifting. Our recent work~\cite{sunil2} is one particular example.
%One way to do this is to determine the set of invertible transforms 
%(or conditions under which a transform is invertible) centrally at the sink. 
%However, these transforms (or conditions) must then be disseminated to the nodes 
%in some manner. This will require some additional overhead to initialize the transform at 
%each node. Moreover, even if nodes were able to determine the invertibility conditions themselves, 
%doing so would still require nodes to exchange transform information with their neighbors.  
%In either case, some additional coordination and communication overhead will be required to 
%ensure that the computations that nodes perform are invertible. 
Given this high degree of complexity to ensure an invertible transform when using broadcast, 
broadcast data should probably only be used with our proposed transforms if 
(i) it is possible to fix the timing in the network in accordance with 
the Definition~\ref{def:neighbors}, and, 
(ii) the timing is very stable.

%AOrev2 Previous section reads much better. It may need a bit of editing for conciseness and to strike a sufficiently positive tone, but otherwise it's a significant improvement over the previous version. 

\section{Unidirectional Transform Designs}
\label{sec:uni}

Proposition~\ref{prop:invertibility} provides simple conditions for invertible transform design, i.e., 
$\Am_n$ is invertible for every node $n$. 
%$\left[ \Am_n \hspace{1mm} \Bm_n^1 \hspace{1mm} \hdots \hspace{1mm} \Bm_n^{|\Bc_b|} \right]$ 
%has L.I. columns for every node $n$. 
This is a simple design constraint that \emph{unifies many existing 
unidirectional transforms}. 
%The main purpose of this section is to demonstrate how 
%existing unidirectional transforms can be mapped into our formulation. 
%The transforms 
%we consider restrict operations to be performed across only neighbors in a tree, in which case, 
%invertibility is easily guaranteed by designing matrices $\Am_n$ which are invertible. 
%Since these transforms are all unidirectional, they will typically be more efficient than process-first transforms. 
In this section, we demonstrate how existing unidirectional transforms 
can be mapped to our formulation. In particular, we focus on 
the tree-based {K}arhunen-{L}o{\`e}ve Transform (T-KLT)~\cite{Shen5}, T-DPCM~\cite{senzip,Shen5} 
and early forms of tree-based wavelet 
transforms~\cite{Lozano,CiancioO:04,CiancioPOK:06,Shen1} constructed using lifting~\cite{Sweldens}. 
%Unidirectional designs on graphs have also been proposed in~\cite{Shen6} by using lifting. 
%We then discuss the various issues related to raw data forwarding in a WSN, then 
%develop a transform design that can minimize the amount of raw data forwarding by 
%applying our framework. This leads to a unidirectional Haar-like wavelet transform with 
%and without broadcasts.
%
%\subsection{Design Considerations}
%\label{sec:uni-design}

In order to exploit spatial correlation to achieve reduction in the number of bits per measurement, 
nodes must first exchange data. Therefore, some nodes must transmit raw data 
to their neighbors before any form of spatial compression can be performed. Since raw 
data typically requires many more bits than encoded transform coefficients, it would be 
desirable to minimize the number of raw data transmissions that nodes must make 
to facilitate distributed transform computation. Therefore, our \emph{main design consideration} 
is to minimize the number of raw data transmissions that are required to compute the transform.

\subsection{Tree-based {K}arhunen-{L}o{\`e}ve Transform}
\label{sec:uni-t-klt}

Since transforms that achieve data decorrelation potentially lead to better coding efficiency~\cite{goyal}, we consider now 
the design of unidirectional transforms that achieve 
the maximum amount of data decorrelation. 
%In this way, only leaf nodes need to forward raw data and the rest transmit only 
%transform coefficients; hence, our main design consideration will be met. 
%In particular, it would be good to minimize the number of bits that each node must transmit to its parent. 
%This will lead to transforms which are most efficient in terms of total cost.
%
This can be achieved by applying, at each node $n$, a transform $\Am_n$ that makes all of the coefficients in 
$\yv_n$ statistically uncorrelated (or ``whitened''), e.g., by using a {K}arhunen-{L}o{\`e}ve transform (KLT) 
at each node, leading to the T-KLT described in our previous work~\cite{Shen5}. 
In this transform, each node $n$ computes and transmits a set of 
``whitened'' coefficients $\yv_n$, which will then have to be ``unwhitened'' and 
then re-whitened at $\rho(n)$ to produce a new set of whitened coefficients. Whitening 
can be done using a KLT and unwhitening can be achieved using an inverse KLT. 
More specifically, this is done at each node $n$ by (i) finding the whitening transform $\Hm_n$ and unwhitening transforms 
of each child $\Gm_{\Cc_n(i)}$, (ii) applying an unwhitening transform to each child to recover the original measurements as 
$\xv_{\Cc_n(i)} = \Gm_{\Cc_n(i)} \cdot \yv_{\Cc_n(i)}$, and then (iii) rewhitening these measurements as 
$\yv_n = \Hm_n \cdot \left[ x(n) \hspace{1mm} \xv_{\Cc_n(1)}^t \hspace{1mm} \hdots \xv_{\Cc_n(|\Cc_n|)}^t \right]^t$. 
This transform (without quantization) can then be expressed in terms of our formulation as 
\begin{equation}
  \yv_n = \Hm_n \cdot 
          \left[ \begin{array}{cccc} 1 & {\bf 0} & \cdots & {\bf 0} \\
                 {\bf 0} & \Gm_{\Cc_n(1)} & \cdots & {\bf 0} \\ 
                 \vdots & \vdots & \ddots & \vdots \\
                 {\bf 0} & {\bf 0} & \cdots & \Gm_{\Cc_n(|\Cc_n|)} \end{array} \right] 
          \cdot 
          \left[ \begin{array}{c} x(n) \\ \yv_{\Cc_n(1)} \\ \vdots \\ \yv_{\Cc_n(|\Cc_n|)} \end{array} \right], 
  \label{eqn:t-klt}
\end{equation}
with $\Am_n = \Hm_n \cdot \text{diag}\left(1, \Gm_{\Cc_n(1)}, \hdots, \Gm_{\Cc_n(|\Cc_n|)}\right)$. 
Each $\Am_n$ is trivially invertible since $\Hm_n$ and each $\Gm_{\Cc_n(i)}$ are invertible 
by construction. Therefore, the tree-based KLT is trivially invertible.

%In general, each $\Hm_n$ and $\Gm_{\Cc_n(i)}$ do not have to be whitening and unwhitening transforms but instead 
%can be anything we like. For instance, each $\Hm_n$ can be some forward wavelet transform on 
%the original data of $n$ and its descendants $\Dc_n$, and each $\Gm_{\Cc_n(i)}$ can represent some 
%inverse wavelet transform. Or if we organize the data from $n$ and $\Dc_n$ onto a 1D line, then 
%$\Hm_n$ and $\Gm_{\Cc_n(i)}$ can be any forward and inverse 1D transforms, respectively. Eq. (\ref{eqn:t-klt}) 
%provides a general form of transforms which are computed by inverting transform coefficients from descendants, 
%then re-transforming all of the raw data together to produce new transform coefficients. 
%We call transforms of this type \emph{encode re-encode transforms}, or \emph{ER-transforms} for short. 
%Choosing these matrices as forward and inverse KLTs just so happens to provide maximal 
%data de-correlation~\cite{Shen5}.

\subsection{Tree-based DPCM}
\label{sec:uni-t-dpcm}

A simpler alternative to the T-KLT is T-DPCM~\cite{senzip,Shen5}. 
A related DPCM based method was proposed in~\cite{pottie}. This particular method is not designed for any 
particular communication structure, but it can easily be adapted to take the form of a unidirectional transform. 
In contrast to the method in~\cite{pottie}, the T-DPCM methods in~\cite{senzip,Shen5} 
compute differentials directly on a tree such as an SPT. 

In the T-DPCM method of~\cite{Shen5}, each node $n$ 
computes its difference with respect to a weighted average of its children's data, i.e., 
$y(n) = x(n) - \sum_{m \in \Cc_n} \av_n(m) x(m)$. For this to be possible, one of two things must 
happen: either every node $n$ must decode the differentials received from its children to 
recover $x(m)$ for each $m \in \Cc_n$, or, every node $n$ must transmit raw data two hops 
forward to its grandparent (at which point $y(n)$ can be computed) to avoid decoding data at every node. 
In order to avoid each node having to forward raw data two hops, at each node $n$, the inverse transform on 
the data of each child $\Cc_n(i)$ must be computed first using the inverse 
matrix $\left(\Am_{\Cc_n(i)}\right)^{-1}$ of each child. The forward transform is then designed accordingly. 
We can express this version of T-DPCM as in (\ref{eqn:t-dpcm-1}).
\begin{equation}
  \yv_n = \left[ \begin{array}{cc} 1 & -\av_n(\Dc_n) \\
                 {\bf 0} & \Id \end{array} \right]
          \cdot 
%          \left[ \begin{array}{cccc} 1 & \Gm_{\Cc_n(1)} & \hdots & \Gm_{\Cc_n(|\Cc_n|)} \\
%                 & \Id & & \\ 
%                 & & \ddots & \\
%                 & & & \Id \end{array} \right] 
          \left[ \begin{array}{cccc} 1 & \left(\Am_{\Cc_n(1)}\right)^{-1} & \cdots & \left(\Am_{\Cc_n(|\Cc_n|)}\right)^{-1} \\
                 {\bf 0} & \Id & \cdots & {\bf 0} \\ 
                 \vdots & \vdots & \ddots & \vdots \\
                 {\bf 0} & {\bf 0} & \cdots & \Id \end{array} \right] 
          \cdot 
          \left[ \begin{array}{c} x(n) \\ \yv_{\Cc_n(1)} \\ \vdots \\ \yv_{\Cc_n(|\Cc_n|)} \end{array} \right]
  \label{eqn:t-dpcm-1}
\end{equation}
The matrix $\Am_n$ is just the product of these triangular matrices, hence, it is trivially invertible. 
Moreover, only leaf nodes need to forward raw data and the rest transmit only 
transform coefficients. 

Alternatively, in the T-DPCM scheme of~\cite{senzip}, 
each node $n$ first forwards raw data $x(n)$ to its parent $\rho(n)$, then 
node $\rho(n)$ computes a differential for $n$ and forwards it to the sink, 
i.e., node $\rho(n)$ computes $y(n) = x(n) - \av_{n}(\rho(n))x(\rho(n))$. 
This transform can also be mapped to our formalism as
%$\text{row}_1(\Am_n) = \left[ -1 \hspace{1mm} 1 \hspace{1mm} {\bf 0} \hspace{1mm} \hdots 
%1 \hspace{1mm} {\bf 0} \right]$ and $\text{row}_j(\Am_n) = \ev_j^t$ for $j \geq 2$ if 
%a simple difference is used. 
\begin{equation}
  \yv_n = \left[ \begin{array}{cc} 1 & {\bf 0} \\
                 -\av_{\Dc_n}(n) & \Id \end{array} \right]
          \cdot 
          \left[ \begin{array}{c} x(n) \\ \yv_{\Dc_n} \end{array} \right].
  \label{eqn:t-dpcm-2}
\end{equation}
This eliminates the computational complexity of the previous T-DPCM method since no decoding of children data is required. 
However, every node must now forward raw data one hop. Moreover, it will not decorrelate the data 
as well as the first method since only data from one neighbor is used.

%\subsection{Unidirectional 5/3-like Wavelets}
%\label{sec:uni-separable-53}
%\subsection{Unidirectional Wavelets}
\subsection{Unidirectional Lifting-based Wavelets}
\label{sec:uni-separable}

We now describe how unidirectional wavelet transforms can be constructed under our 
framework. This can be done using lifting~\cite{Sweldens}. 
Lifting transforms are constructed by 
\emph{splitting} nodes into disjoint sets of \emph{even} and \emph{odd} nodes, by 
designing \emph{prediction filters}, which alter odd data using even data, and 
\emph{update filters}, which alter even data based on 
odd data. They are invertible by construction~\cite{Sweldens}. 

\begin{figure}[htb]
 \centerline{\psfig{figure=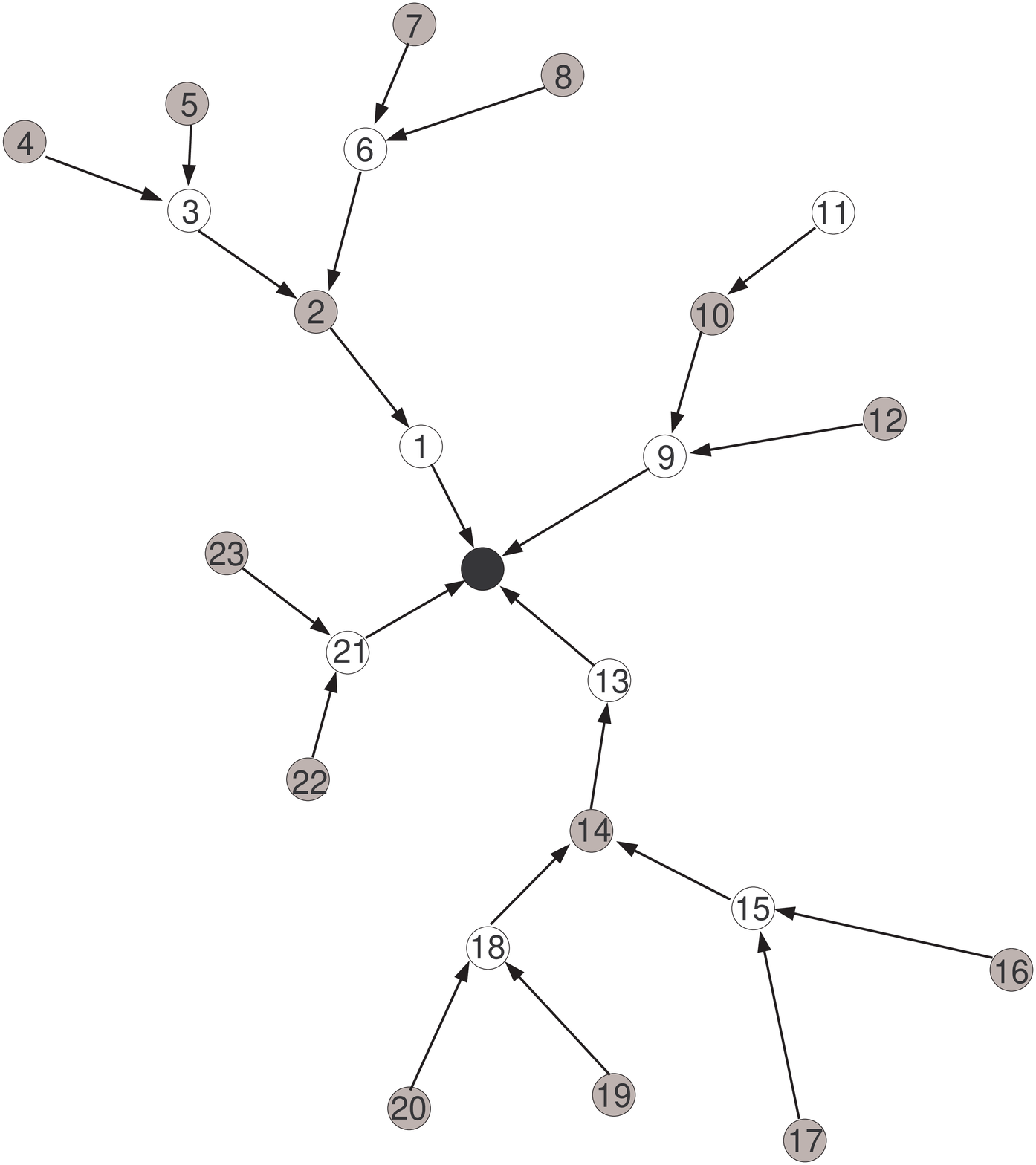,width=6.5cm,height=6.5cm}}
  \caption{{\footnotesize Example of splitting based on the depth of the routing tree. White (odd depth) nodes are odd, 
  gray (even depth) nodes are even and the black center node is the sink.}}
\label{fig:SplitExample}
\end{figure}

%These transforms are constructed on a tree $T$ as follows. 
%First, nodes are split into odd and even sets $\Oc$ and $\Ec$, respectively, 
%by assigning nodes of odd depth as odd and nodes of even depth as even. More specifically, 
%$\Oc = \{ n : h(n) \mod 2 = 1 \}$ and $\Ec = \{ m : h(m) \mod 2 = 0 \}$. 
%This is illustrated in Fig.~\ref{fig:SplitExample}. 
First, nodes are split into odd and even sets $\Oc$ and $\Ec$, respectively. 
This can be done completely arbitrarily. One example from our previous work~\cite{Shen1} 
is to split according to the depth in the tree, e.g., 
$\Oc = \{ n : h(n) \mod 2 = 1 \}$ and $\Ec = \{ m : h(m) \mod 2 = 0 \}$, as 
illustrated in Fig.~\ref{fig:SplitExample}. 
%Data at each odd node $n \in \Oc$ is then predicted using data from even neighbors in $T$ 
%Data at each odd node $n \in \Oc$ is then predicted using data from even neighbors in $T$ 
%and potentially data from even broadcast neighbors, i.e., 
Data at each odd node $n \in \Oc$ is then predicted using data from even neighbors 
$\Nc_n \subset \Ec$, yielding detail coefficient
%$\Nc_n \subset \Ec \cap (\Dc_n \cup \Ac_n)$ containing only even neighbors which reside 
%on the same subtree as $n$, yielding detail coefficient
\begin{equation}
\label{eq:predict-tree}
%d(n) = x(n) + \sum_{i \in \{\rho(n)\} \cup \Cc_n} \pv_{n}(i) x(i).
d(n) = x(n) - \sum_{i \in \Nc_n} \pv_{n}(i) x(i).
\end{equation}
The prediction vector $\pv_{n}$ can provide a simple average~\cite{Shen1}, 
i.e., $\pv_{n}(i) = \frac{1}{|\Nc_n|}$ for each $i \in \Nc_n$, 
%i.e., $\pv_{n}(i) = \frac{1}{1+|\Cc_n|}$ for each $i \in \{\rho(n)\} \cup \Cc_n$, 
a planar prediction~\cite{Wagner2} of the data at node $n$ using data 
from its neighbors, or can even be data adaptive~\cite{Shen5}. 
Incorporating some broadcast data into the prediction is also useful since 
it allows odd nodes to achieve even further decorrelation. 
%After the prediction step, data at each even node $m \in \Ec$ is updated using details from odd neighbors in $T$, i.e., 
%$\Nc_m \subset \Oc$, 
After the prediction step, data at each even node $m \in \Ec$ is updated using details from odd neighbors 
$\Nc_m \subset \Oc$, 
yielding smooth coefficient
\begin{equation}
\label{eq:update-tree}
%s(m) = x(m) + \sum_{j \in \{\rho(m)\} \cup \Cc_m} \uv_{m}(j) d(j).
s(m) = x(m) + \sum_{j \in \Nc_m} \uv_{m}(j) d(j).
\end{equation}
The update vector $\uv_m$ can provide simple smoothing~\cite{Shen1}, i.e., 
%$\uv_m(j) = \frac{1}{2\cdot(1+|\Cc_n|)}$ for all $j \in \rho(m) \cup \Cc_m$, or could 
$\uv_m(j) = \frac{1}{2\cdot|\Nc_m|}$ for all $j \in \Nc_m$, or could 
provide orthogonality between smooth (i.e., low-pass) and 
detail (i.e., high-pass), coefficients~\cite{Shen6}. 
%In this work, we use the prediction design in~\cite{Wagner1} and the 
%update design in~\cite{Shen6} as they provide more accurate predictions 
%and greater energy compaction. 

By the lifting construction, invertibility will be guaranteed as long as 
(i) odd node data is only predicted using even node data, and (ii) 
even node data is only updated using details from odd nodes. So if $\Ec$ and $\Oc$ is 
an arbitrary even and odd split, the transform computed at each node 
will be invertible as long as the computations satisfy (i) and (ii). 
More formally, let $\Oc_n = (n \cup \Dc_n) \cap \Oc$ be the set of 
odd nodes whose data is available at $n$ from its subtree. Let 
$\Ec_n = (n \cup \Dc_n) \cap \Ec$ be defined similarly. Moreover, let 
$\Oc_n^{\Bc} = \bar{\Bc}_n \cap \Oc$ denote the set of odd nodes whose 
data $n$ receives via broadcast. Similarly, let $\Ec_n^{\Bc} = \bar{\Bc}_n \cap \Ec$. 
Then the computations at $n$ will be invertible as long as 
it only predicts $y(\Oc_n)$ from $y(\Ec_n)$ and $y(\Ec_n^{\Bc})$ and 
only updates $y(\Ec_n)$ from $y(\Oc_n)$ and $y(\Oc_n^{\Bc})$. 
Let $\Mm_n$ and $\Mm_n^{\Bc}$ be permutation matrices such that 
\begin{equation}
\left[\begin{array}{c} y(\Oc_n) \\ y(\Ec_n) \\ y(\Oc_n^{\Bc}) \\ y(\Ec_n^{\Bc}) \end{array}\right] = 
     \left[\begin{array}{cc} \Mm_n & {\bf 0} \\ {\bf 0} & \Mm_n^{\Bc} \end{array}\right] 
     \cdot 
     \left[\begin{array}{c} x(n) \\ y(\Dc_n) \\ y(\bar{\Bc}_n) \end{array}\right].
  \label{eqn:lifting-relabeling}
\end{equation}
Then $n$ can compute transform coefficients as in (\ref{eqn:local-lifting}). 
\begin{equation}
\yv_n = \left(\Mm_n\right)^t
     %\cdot
     \left[\begin{array}{cccc} 
         \Id & {\bf 0} & {\bf 0} & {\bf 0} \\
         \Um_n & \Id & \Um_n^{\Bc} & {\bf 0}
     \end{array}\right] 
     %\cdot
     \left[\begin{array}{cccc} 
         \Id & \Pm_n & {\bf 0} & \Pm_n^{\Bc} \\
         {\bf 0} & \Id & {\bf 0} & {\bf 0} \\
         {\bf 0} & {\bf 0} & \Id & {\bf 0} \\
         {\bf 0} & {\bf 0} & {\bf 0} & \Id
     \end{array}\right] 
     %\cdot
     \left[\begin{array}{c} y(\Oc_n) \\ y(\Ec_n) \\ y(\Oc_n^{\Bc}) \\ y(\Ec_n^{\Bc}) \end{array}\right]
%     \left[\begin{array}{cc} \Mm_n & {\bf 0} \\ {\bf 0} & \Mm_n^{\Bc} \end{array}\right] 
%     %\cdot 
%     \left[\begin{array}{c} x(n) \\ y(\Dc_n) \\ y(\bar{\Bc}_n) \end{array}\right].
  \label{eqn:local-lifting}
\end{equation}
By multiplying the matrices in (\ref{eqn:lifting-relabeling}) and (\ref{eqn:local-lifting}) together, we get 
$\yv_n = \left[\Am_n \hspace{1mm} \Bm_n\right] \cdot \left[x(n) \hspace{1mm} \yv_{\Dc_n}^t \hspace{1mm} \yv_{\Bc_n}^t\right]^t$, 
with 
\begin{equation*}
\Am_n = \left(\Mm_n\right)^t \cdot \left[\begin{array}{cc} \Id & {\bf 0} \\ \Um_n & \Id \end{array}\right] \cdot
        \left[\begin{array}{cc} \Id & \Pm_n \\ {\bf 0} & \Id \end{array}\right] \cdot \Mm_n, 
\end{equation*}
%and 
\begin{equation*}
\Bm_n = \left(\Mm_n\right)^t \cdot \left[\begin{array}{cc} {\bf 0} & \Pm_n^{\Bc} \\ \Um_n^{\Bc} & \Um_n \Pm_n^{\Bc} \end{array}\right] \cdot \Mm_n^{\Bc}.
\end{equation*}
Since $\det(\Am_n) = 1$, single-level unidirectional lifting transforms are always invertible.
%Given our main design consideration (i.e., to minimize the number of raw data transmissions), it would be best to 
%minimize the number of nodes that must transmit raw data and also perform as many levels 
%of transform decomposition as possible at each node.

The transform given by (\ref{eqn:local-lifting}) corresponds to only one level of 
decomposition. In particular, at each node $n$ the transform of (\ref{eqn:local-lifting}) will yield a set 
of smooth (or low-pass) coefficients $\{y(k)\}_{k \in \Ec_n}$ and a set of detail (or high-pass) 
coefficients $\{y(l)\}_{l \in \Oc_n}$. The high-pass coefficients 
%AO3 I would be careful about claiming too much. Generally, high pass information will have low energy if signals are smooth, and so will require relatively few bits. Beyond that, it's not clear that we can claim that they are uncorrelated (spatially?). Conversely, it is safe to say that for smooth signals the update signal has high spatial correlation. 
will typically have low energy if the original data is smooth, so these 
can be encoded using very few bits and forwarded to the sink without any further processing. However, there 
will still be some correlation between low-pass coefficients. It would therefore be useful to apply additional 
levels of transform to the low-pass coefficients at node $n$ to achieve more decorrelation. This will reduce 
the number of bits needed to encode these low-pass coefficients, and will ultimately reduce the number of bits 
each node must transmit to the sink. 
%GSrev1 This sentence is repetitive
%In particular, we can apply additional lifting transforms to the low-pass coefficients.

Suppose each node performs an additional $J$ levels of lifting transform on the low-pass coefficients $\{y(k)\}_{k \in \Ec_n}$. 
%For the first additional level suppose that we split nodes in $\Ec_n$ into even and odd 
%sets $\Ec_n^2$ and $\Oc_n^2$, respectively. Then for each odd node $l \in \Oc_n^2$ we can predict $y(l)$ using even coefficients 
%$y(k)$ for some set of even neighbors $\Nc_l^2 \subset \Ec_n^2$, i.e., 
%$y(l) = y(l) + \sum_{k \in \Nc_l^2} \pv_{l,2}(k) y(k)$. Then for each even node $k \in \Ec_n^2$ we can update 
%$y(k)$ using odd coefficients $y(l)$ for some set of odd neighbors $\Nc_k^2 \subset \Oc_n^2$, i.e., 
%$y(k) = y(k) + \sum_{l \in \Nc_k^2} \uv_{k,2}(l) y(l)$. 
%More generally, at each level $j = 2, 3, \hdots, J+1$, suppose that we split nodes in 
At each level $j = 2, 3, \hdots, J+1$, suppose that nodes in 
$\Ec_n^{j-1}$ are split into even and odd sets $\Ec_n^j$ and $\Oc_n^j$, respectively. We assume 
that $\Ec_n^1 = \Ec_n$. 
For each odd node $l \in \Oc_n^j$, we predict $y(l)$ using even coefficients 
from some set of even neighbors $\Nc_l^j \subset \Ec_n^j$, i.e., 
$y(l) = y(l) - \sum_{k \in \Nc_l^j} \pv_{l,j}(k) y(k)$. Then for each even node $k \in \Ec_n^j$, we update 
$y(k)$ using odd coefficients from some set of odd neighbors $\Nc_k^j \subset \Oc_n^j$, i.e., 
$y(k) = y(k) + \sum_{l \in \Nc_k^j} \uv_{k,j}(l) y(l)$. This decomposition is done starting from 
level $j = 2$ up to level $j = J+1$. For all $j = 2, 3, \hdots, J+1$, let $\Mm_n^j$ be a 
permutation matrix such that 
\begin{equation}
\left[\begin{array}{c} y(\Oc_n^j) \\ y(\Ec_n^j) \\ y(\Rc_n^j) \end{array}\right] = 
     \Mm_n^j \cdot \yv_n, 
  \label{eqn:lifting-relabeling-j}
\end{equation}
where $\Rc_n^j = (n \cup \Dc_n) - (\Oc_n^j \cup \Ec_n^j)$ is the set of nodes whose coefficients 
are not modified at level $j$. Then we can express the level $j$ transform computations in matrix form as 
\begin{equation}
\yv_n = \left(\Mm_n^j\right)^t
     \cdot
     \left[\begin{array}{ccc} 
         \Id & {\bf 0} & {\bf 0} \\
         \Um_n^j & \Id & {\bf 0} \\
         %\Um_{n,j} & \Id & {\bf 0} \\
         {\bf 0} & {\bf 0} & \Id
     \end{array}\right] 
     \cdot
     \left[\begin{array}{ccc} 
         \Id & \Pm_n^j & {\bf 0} \\
         %\Id & \Pm_{n,j} & {\bf 0} \\
         {\bf 0} & \Id & {\bf 0} \\
         {\bf 0} & {\bf 0} & \Id
     \end{array}\right] 
     \cdot
%     \Mm_n^j
%     \cdot
%     \yv_n.
     \left[\begin{array}{c} y(\Oc_n^j) \\ y(\Ec_n^j) \\ y(\Rc_n^j) \end{array}\right],
%     \left[\begin{array}{cc} \Mm_n & {\bf 0} \\ {\bf 0} & \Mm_n^{\Bc} \end{array}\right] 
%     %\cdot 
%     \left[\begin{array}{c} x(n) \\ y(\Dc_n) \\ y(\bar{\Bc}_n) \end{array}\right].
  \label{eqn:local-lifting-j}
\end{equation}
%GSrev1 Clarify notation used here
where $\Pm_n^j$ and $\Um_n^j$ represent the prediction and update operations used at level $j$, respectively.

By combining (\ref{eqn:lifting-relabeling}), (\ref{eqn:local-lifting}), (\ref{eqn:lifting-relabeling-j}) 
and (\ref{eqn:local-lifting-j}), we finally get that 
$\yv_n = \left[\Am_n \hspace{1mm} \Bm_n\right] \cdot \left[x(n) \hspace{1mm} \yv_{\Dc_n}^t \hspace{1mm} \yv_{\Bc_n}^t\right]^t$, 
with $\Am_n$ and $\Bm_n$ defined in (\ref{eqn:multilevel-A}) and (\ref{eqn:multilevel-B}). 
\begin{equation}
\Am_n = \prod_{j = 2}^{J+1} \left(\left(\Mm_n^j\right)^t
%     \cdot
     \left[\begin{array}{ccc} 
         \Id & {\bf 0} & {\bf 0} \\
         \Um_n^j & \Id & {\bf 0} \\
         %\Um_{n,j} & \Id & {\bf 0} \\
         {\bf 0} & {\bf 0} & \Id
     \end{array}\right] 
%     \cdot
     \left[\begin{array}{ccc} 
         \Id & \Pm_n^j & {\bf 0} \\
         %\Id & \Pm_{n,j} & {\bf 0} \\
         {\bf 0} & \Id & {\bf 0} \\
         {\bf 0} & {\bf 0} & \Id
     \end{array}\right] 
%     \cdot
     \Mm_n^j \right)
%     \cdot
     \left(\Mm_n\right)^t
%     \cdot 
     \left[\begin{array}{cc} \Id & {\bf 0} \\ \Um_n & \Id \end{array}\right] 
%     \cdot 
     \left[\begin{array}{cc} \Id & \Pm_n \\ {\bf 0} & \Id \end{array}\right] 
%     \cdot 
     \Mm_n
  \label{eqn:multilevel-A}
\end{equation}
\begin{equation}
\Bm_n = \prod_{j = 2}^{J+1} \left(\left(\Mm_n^j\right)^t
%     \cdot
     \left[\begin{array}{ccc} 
         \Id & {\bf 0} & {\bf 0} \\
         \Um_n^j & \Id & {\bf 0} \\
         %\Um_{n,j} & \Id & {\bf 0} \\
         {\bf 0} & {\bf 0} & \Id
     \end{array}\right] 
%     \cdot
     \left[\begin{array}{ccc} 
         \Id & \Pm_n^j & {\bf 0} \\
         %\Id & \Pm_{n,j} & {\bf 0} \\
         {\bf 0} & \Id & {\bf 0} \\
         {\bf 0} & {\bf 0} & \Id
     \end{array}\right] 
%     \cdot
     \Mm_n^j \right)
%     \cdot
     \left(\Mm_n\right)^t 
%     \cdot 
     \left[\begin{array}{cc} {\bf 0} & \Pm_n^{\Bc} \\ \Um_n^{\Bc} & \Um_n \Pm_n^{\Bc} \end{array}\right] 
%     \cdot 
     \Mm_n^{\Bc}
  \label{eqn:multilevel-B}
\end{equation}
Prop.~\ref{prop:invertibility} implies that the overall transform is invertible if 
$\Am_n$ given in (\ref{eqn:multilevel-A}) is invertible. Since each $\Mm_n^j$ is a permutation matrix, 
$|\det(\Mm_n^j)| = 1$. Moreover, the remaining matrices are triangular. 
Thus, it easily follows that $\det(\Am_n) = 1$. 
%Since the determinant of $\Am_n$ is simply the product of the determinants of the matrices 
%comprising $\Am_n$, it easily follows that $\det(\Am_n) = 1$. 
Therefore, unidirectional, multi-level lifting transforms are always invertible.

\subsection{Unidirectional 5/3-like Wavelets}
\label{sec:uni-separable-53}

\begin{figure}[htb]
 \centerline{\psfig{figure=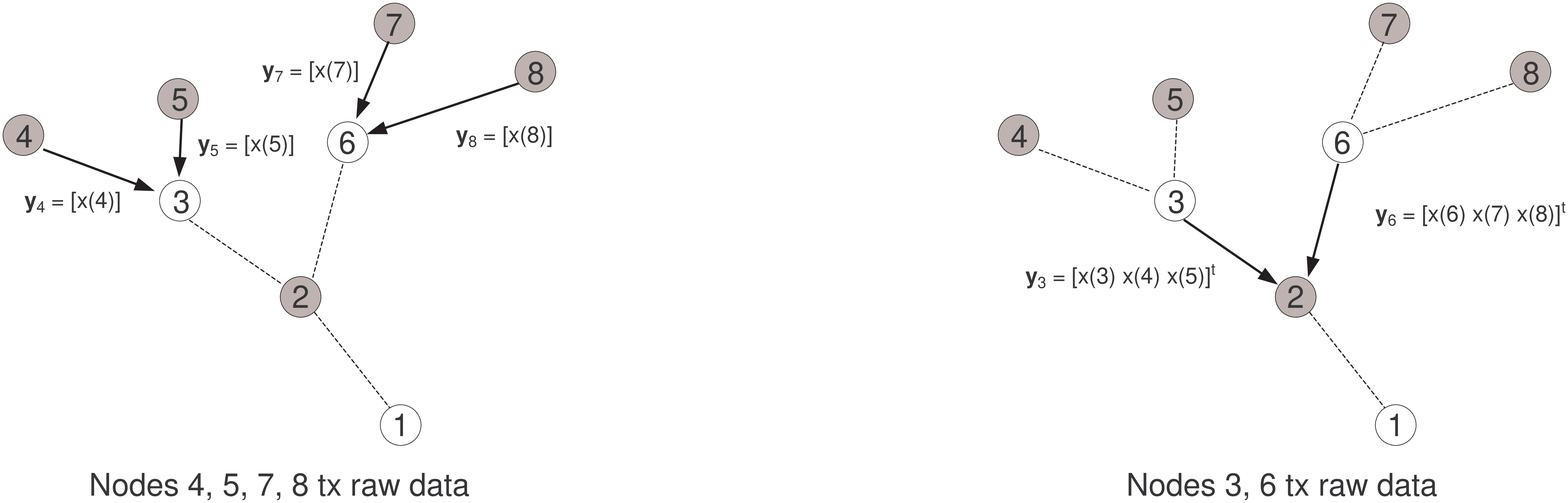,width=12cm,height=5cm}}
  \caption{{\footnotesize Raw data transmissions for 5/3-like transform. Nodes 3 and 6 need 
  $x(2)$ to compute details $d(3)$ and $d(6)$, so they must forward raw data over 1-hop to node 2. 
  Nodes 4 and 5 need $d(3)$ to compute $s(4)$ and $s(5)$, so they must forward raw data over 2-hops.}}
\label{fig:TransmissionExample_53}
\end{figure}

Our previous work~\cite{Shen1} provides a \emph{5/3-like transform} on a tree. 
First, nodes are split into odd and even sets $\Oc$ and $\Ec$, respectively, 
by assigning nodes of odd depth as odd and nodes of even depth as even. More specifically, 
$\Oc = \{ n : h(n) \mod 2 = 1 \}$ and $\Ec = \{ m : h(m) \mod 2 = 0 \}$. 
This is illustrated in Fig.~\ref{fig:SplitExample}. 
The transform neighbors of each node are simply $\Nc_n = \{\rho(n)\} \cup \Cc_n$ for every node $n$. 
This provides a \emph{5/3-like} wavelet transform on a tree since whenever averaging predictions and smoothing 
updates are used along a 1D path, the transform reduces to the 5/3 wavelet transform~\cite{mallat}. 
Nodes can compute these transforms in a unidirectional manner, but doing so requires that some nodes forward 
raw data 1 or 2 hops. This is illustrated in Fig.~\ref{fig:TransmissionExample_53}. 

Data from each odd node $n$ is predicted using data $x(\Cc_n)$ (from children $\Cc_n$) and 
$x(\rho(n))$ (from parent $\rho(n)$). 
However, odd node $n$ will not have $x(\rho(n))$ locally available for processing. 
Therefore, we require that each odd node $n$ transmit 
raw data $x(n)$ one hop forward to its parent $\rho(n)$, at which point node $\rho(n)$ can compute the 
detail coefficient of $n$. Each even node $m$ will then compute detail 
$d(j) = x(j) - \sum_{i \in \Cc_j} \pv_l(j) x(j) - \pv_j(m) x(m)$ for every child $j \in \Cc_m$. 
Similarly, the smooth coefficient of each even 
node $m$ requires details from its parent $\rho(m)$ and children $\Cc_m$, 
so it can not be locally computed either. Moreover, detail $d(\rho(m))$ can 
only be computed at node $\rho^2(m)$, i.e., at the grandparent of $m$. Therefore, we require that even 
node $m$ transmit raw data $x(m)$ two hops forward to $\rho^2(m)$, at which point 
$d(\rho(m))$ will be available and $\rho^2(m)$ can compute 
$s(m) = x(m) + \sum_{j \in \{\rho(m)\} \cup \Cc_m} \uv_m(j) d(j)$. 
%In particular, every even node $m$ computes 
%$s(k) = x(k) + \sum_{j \in \{\rho(k)\} \cup \Cc_k} \uv_k(j) d(j)$ for all $k \in \Cc_m^2$. 
Note that each of these operations are trivially invertible, and easily lead to 
local transform matrices $\Am_n$ which are invertible by construction. 
However, the number of raw data transmissions is relatively high, i.e., 
1-hop for odd nodes and 2-hops for even nodes. 
%This results in a large number of 
%raw data transmissions, leading to a transform which is rather inefficient. 
%We eliminate most of these raw data transmissions in the following section.
We address this inefficiency in the next section.

\section{Unidirectional Haar-like Wavelets}
%\subsection{Unidirectional Haar-like Wavelets}
\label{sec:uni-separable-haar}

%AO3 the beginning of this section repeats the end of the previous section, it could be shortened, so concepts are no repeated. 
For the transform in Section~\ref{sec:uni-separable-53}, 
raw data from even and odd nodes must be forwarded over 2-hops and 1-hop, respectively. 
This can be inefficient in terms of transport costs. 
Instead, it would be better to construct a lifting transform which 
directly minimizes the number of raw data transmissions each node must make. 
We use the splitting method in Section~\ref{sec:uni-separable-53}. 
Note that some form of data exchange must occur before the transform can be computed, 
%AOrev1 Corrected. I assume this was what you were trying to say
%i.e., evens must raw transmit data to odds, or viceversa. 
i.e., evens must transmit raw data to odds, or viceversa. 
Suppose that even nodes forward raw data to their parents. 
%So in light of our main design consideration, 
In this case, the best we can do is to design a transform for which 
even nodes transmit raw data over only 1-hop, and odd nodes do not 
transmit any raw data. This will minimize the number of 
%transmit raw data (unless it is impossible). This will minimize the number of 
raw data transmissions that nodes need to make, leading to transforms 
which are more efficient than the 5/3-like transform in terms of transport costs. 
We note that minimizing raw data only serves as a simple proxy for the optimization. 
A more formal optimization which relies on this same intuition 
is undertaken in our recent work~\cite{sunil2}.

\subsection{Transform Construction}
\label{sec:haar-construction}

%AO3 More efficient than what? 
A design that is more efficient than the 5/3-like transform can be achieved as follows. 
Note that an odd node $n$ 
has data from its children $\Cc_n$ and/or even broadcast neighbors 
$\Bc_n \cap \Ec$ locally available, so it can directly compute 
a detail coefficient for itself, i.e., 
$d(n) = x(n) - \sum_{i \in \Cc_n} \pv_n(i) x(i) - \sum_{j \in \Bc_n \cap \Ec} \pv_n(j) x(j)$.
%$d(n) = x(n) - \sum_{i \in \Nc_n} \pv_n(i) x(i)$ 
%with $\Nc_n = \Cc_n \cup (\Bc_n \cap \Ec)$. 
Thus, the detail $d(n)$ is computed directly at $n$, is 
encoded, and then is transmitted to the sink. 
These details require fewer bits for encoding than raw data, 
hence, this reduces the number of bits that odd nodes must transmit for their own data. 
Since data from even node $m$ is only used to predict data at its parent 
$\rho(m)$, we simply have that $\Nc_m = \{\rho(m)\}$ and 
$s(m) = x(m) + \uv_m(\rho(m)) d(\rho(m))$. Moreover, these 
smooth coefficients can be computed at each odd node $n$. Therefore, 
even nodes only need to forward raw data over one hop, after which their 
smooth coefficients can be computed. 
Note that not all odd nodes will have children or even broadcast neighbors, i.e., 
there may exist some odd nodes $n$ such that $\Cc_n = \emptyset$ and $\Bc_n \cap \Ec = \emptyset$. 
Such odd nodes can simply forward raw data $x(n)$ to their parent $\rho(n)$, 
then $\rho(n)$ can compute their details as $d(n) = x(n) - \pv_n(\rho(n)) x(\rho(n))$. 
Thus, there may be a few odd nodes that must send raw data forward one hop. 
This leads to a \emph{Haar-like transform} which is exactly the Haar wavelet transform when 
applied to 1D paths.

\begin{figure}[htb]
 \centerline{\subfigure[$1^{st}$-level along $T$]{\includegraphics[width=8cm,height=4cm]{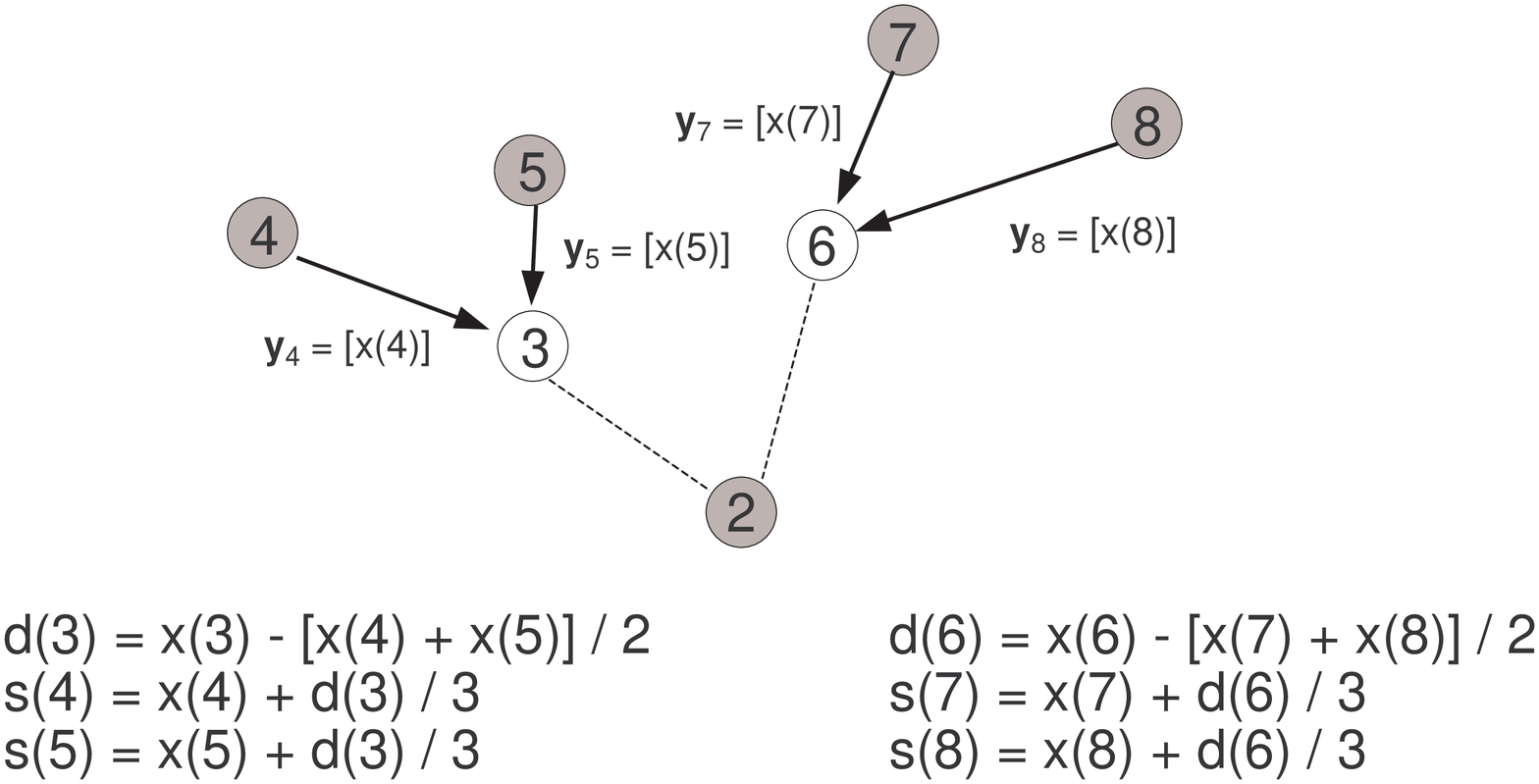}
 \label{fig:HaarExample_a}}
 \subfigure[$2^{nd}$-level ``orthogonal'' to $T$]{\includegraphics[width=8cm,height=4cm]{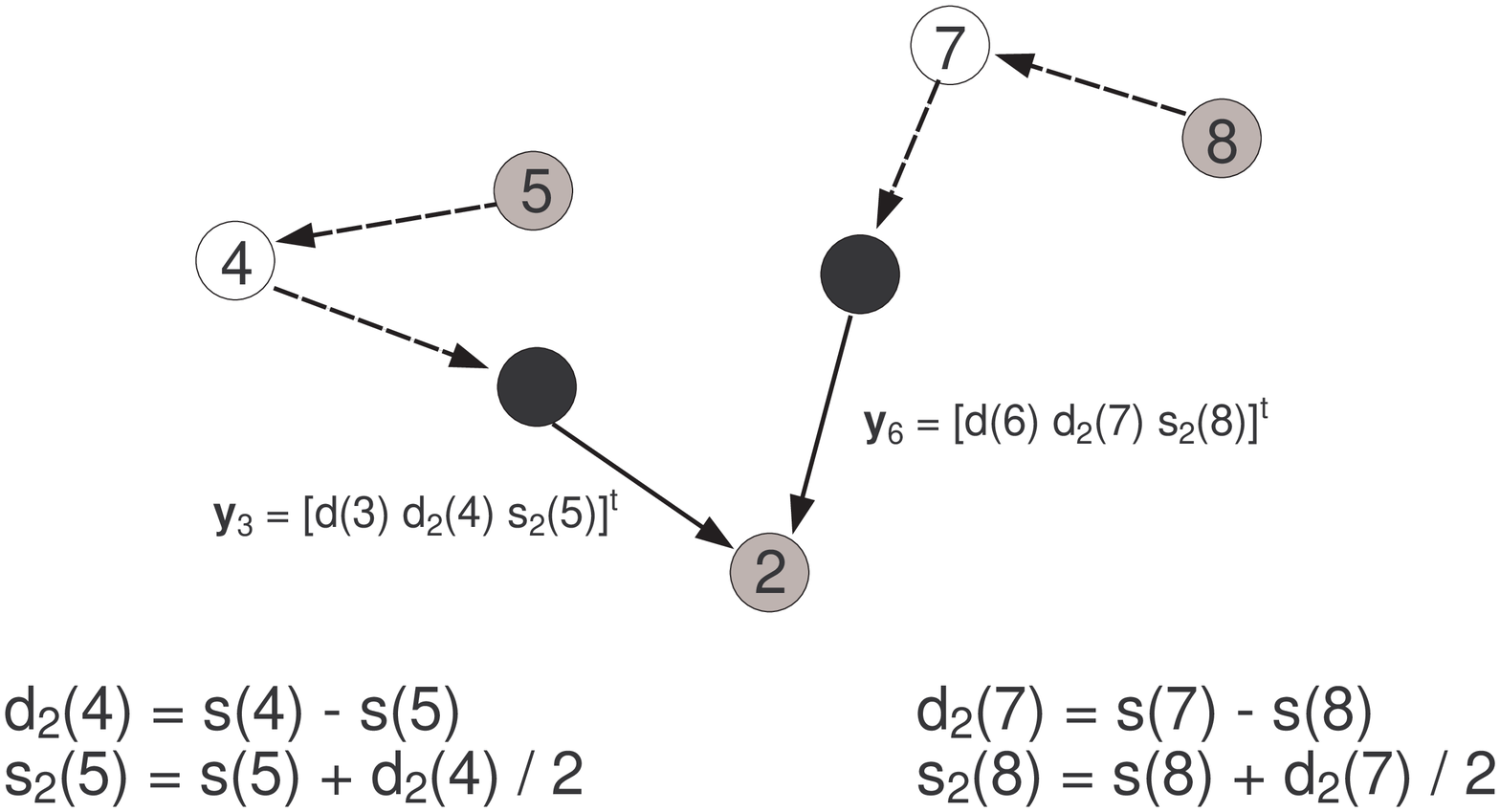}
 \label{fig:HaarExample_b}}}
 \caption{{\footnotesize Unidirectional Computations for Haar-like Transform. In (a), nodes 3 and 6 compute a 
 first level of transform. Then in (b), nodes 3 and 6 compute a 
 second level of transform on smooth coefficients of their children.}}
 \label{fig:TransmissionExample_Haar}
\end{figure}

Odd nodes can also perform additional levels of decomposition on the smooth coefficients of 
their descendants. In particular, every odd node $n$ will 
locally compute the smooth coefficients of its children. Therefore, it can organize 
the smooth coefficients $\{s(k)\}_{k \in \Cc_n}$ onto another tree $T_n^2$ and perform more levels 
of transform decomposition along $T_n^2$. In this work, we assume $T_n^2$ is a minimum spanning tree. 
This produces detail coefficients $\{d_2(k)\}_{k \in \Oc_n^2}$, $\{d_3(k)\}_{k \in \Oc_n^3}$, 
$\ldots$, $\{d_{J+1}(k)\}_{k \in \Oc_n^{J+1}}$ and smooth coefficients $\{s_{J+1}(k)\}_{k \in \Ec_n^{J+1}}$ for some 
$J \geq 0$. In this way, odd nodes can further decorrelate the 
data of their children before they even transmit. This reduces the resources they consume in transmitting 
data. 
An example of this separable transform for $J = 1$ is illustrated in Fig.~\ref{fig:TransmissionExample_Haar}. 
By choosing averaging prediction filters and the orthogonalizing update filter design in~\cite{Shen6}, we get 
the global equation in (\ref{eqn:haar-example}).
\begin{equation}
\yv_3 = \left[ \begin{array}{ccc} 1 & 0 & 0 \\ 0 & 1 & 0 \\ 0 & \frac{1}{2} & 1 \end{array} \right] 
        \cdot 
        \left[ \begin{array}{ccc} 1 & 0 & 0 \\ 0 & 1 & -1 \\ 0 & 0 & 1 \end{array} \right] 
        \cdot 
        \left[ \begin{array}{ccc} 1 & 0 & 0 \\ \frac{1}{3} & 1 & 0 \\ \frac{1}{3} & 0 & 1 \end{array} \right] 
        \cdot 
        \left[ \begin{array}{ccc} 1 & -\frac{1}{2} & -\frac{1}{2} \\ 0 & 1 & 0 \\ 0 & 0 & 1 \end{array} \right] 
        \left[ \begin{array}{c} x(3) \\ x(4) \\ x(5) \end{array} \right]
 \label{eqn:haar-example}
\end{equation}
The coefficient vector $\yv_6$ is obtained in a similar manner. 
More generally, these sorts of multi-level transform computations can always be formulated into matrices as described 
in Section~\ref{sec:uni-separable}. 

\subsection{Discussion}

%AO3 the previous subsection explained the filters based on some examples, so this section is important since it provides the more general formulation. In that sense it may be useful to include definitions that are general, and include all cases. What does not seem to be captured here is what happens in cases where nodes do not have children, etc. Would it be useful to include this here? or does the notation already encompass those cases? 
The transform computations that each node performs can be easily mapped into 
our standard form 
$\yv_n = \left[\Am_n \hspace{1mm} \Bm_n\right] \cdot \left[x(n) \hspace{1mm} \yv_{\Dc_n}^t \hspace{1mm} \yv_{\Bc_n}^t\right]^t$
by appropriately populating the matrices in (\ref{eqn:multilevel-A}) and (\ref{eqn:multilevel-B}). 
Therefore, they will always yield invertible transforms. 
For example, since each odd node $n$ predicts its own data $x(n)$ using data from its children $\Cc_n$ and even 
broadcast neighbors $\Bc_n \cap \Ec$, then updates the data of 
its children from its own detail, the operations for a single level transform at odd $n$ can be expressed as 
\begin{equation}
  \yv_n = \left[ \begin{array}{cc} 1 & {\bf 0} \\
                        \uv_{\Dc_n}(n) & \Id \end{array} \right] 
                 %\cdot 
                 \left[ \begin{array}{ccc} 1 & -\pv_n(\Dc_n) & -\pv_n(\bar{\Bc}_n) \\
                        {\bf 0} & \Id & {\bf 0} \end{array} \right] 
                 %\cdot 
                 \left[ \begin{array}{c} x(n) \\ \yv_{\Dc_n} \\ \yv_{\Bc_n} \end{array} \right]. 
  \label{eqn:haar-odd-equation}
\end{equation}
By choosing 
\begin{equation}
  \Am_n = \left[ \begin{array}{cc} 1 & {\bf 0} \\
                        \uv_{\Dc_n}(n) & \Id \end{array} \right] 
                 \cdot \left[ \begin{array}{cc} 1 & -\pv_n(\Dc_n) \\
                        {\bf 0} & \Id \end{array} \right], 
  \label{eqn:haar-odd-matrix-A}
\end{equation}
and 
\begin{equation}
  \Bm_n = \left[ \begin{array}{cc} 1 & {\bf 0} \\
                        \uv_{\Dc_n}(n) & \Id \end{array} \right] 
                 \cdot \left[ \begin{array}{c} -\pv_n(\bar{\Bc}_n) \\
                        \Id \end{array} \right], 
  \label{eqn:haar-odd-matrix-B}
\end{equation}
we have that 
$\yv_n = [\Am_n \hspace{1mm} \Bm_n] \cdot \left[ x(n) \hspace{1mm} \yv_{\Dc_n}^t \hspace{1mm} \yv_{\Bc_n}^t \right]^t$. 
Note that (\ref{eqn:haar-odd-equation}) covers all of the cases discussed in Section~\ref{sec:haar-construction} 
for each odd node $n$, that is to say: (i) 
$\Cc_n \neq \emptyset$ and $\Bc_n \cap \Ec \neq \emptyset$, (ii) $\Cc_n = \emptyset$ and $\Bc_n \cap \Ec \neq \emptyset$, 
(iii) $\Cc_n \neq \emptyset$ and $\Bc_n \cap \Ec = \emptyset$, and 
(iv) $\Cc_n = \emptyset$ and $\Bc_n \cap \Ec = \emptyset$. In particular, whenever 
$\Cc_n \neq \emptyset$, $\pv_n(\Dc_n)$ and $\uv_{\Dc_n}(n)$ will have some non-zero entries. Otherwise, 
$n$ has no descendants and so 
$\pv_n(\Dc_n)$ and $\uv_{\Dc_n}(n)$ will just be vectors of zeros. Similarly, whenever 
$\Bc_n \cap \Ec \neq \emptyset$, $\pv_n(\bar{\Bc}_n)$ will have some non-zero entries. Otherwise, 
$n$ has no even broadcast neighbors and $\pv_n(\bar{\Bc}_n)$ will be a vector of zeros.

Similarly, each even node $m$ may need to compute predictions for its odd children, so its computations 
for a single level transform can be expressed as
\begin{equation}
  \yv_m = \left[ \begin{array}{ccc} 1 & {\bf 0} \\
                        -\pv_{\Dc_m}(m) & \Id \end{array} \right] 
                 \cdot 
         \left[ \begin{array}{c} x(m) \\ \yv_{\Dc_n}  \end{array} \right]. 
  \label{eqn:haar-even-predict}
\end{equation}
Also note that (\ref{eqn:haar-even-predict}) covers all of the cases for each even node $m$ 
discussed in Section~\ref{sec:haar-construction}, i.e., 
when $m$ has to compute predictions for children then $\pv_{\Dc_m}(m) \neq {\bf 0}$, otherwise, 
$\pv_{\Dc_m}(m) = {\bf 0}$.
%when $m$ has to compute predictions for children then there will be some non-zero entries in 
%$\pv_{\Dc_m}(m)$, otherwise, $\pv_{\Dc_m}(m)$ is just a vector of zeros.

\begin{figure}[htb]
 %\centerline{\subfigure[Without Broadcasts]{\includegraphics[width=5cm,height=5cm]{Broadcast1_a.eps}
 \centerline{\subfigure[Without Broadcasts]{\includegraphics[width=4cm,height=4cm]{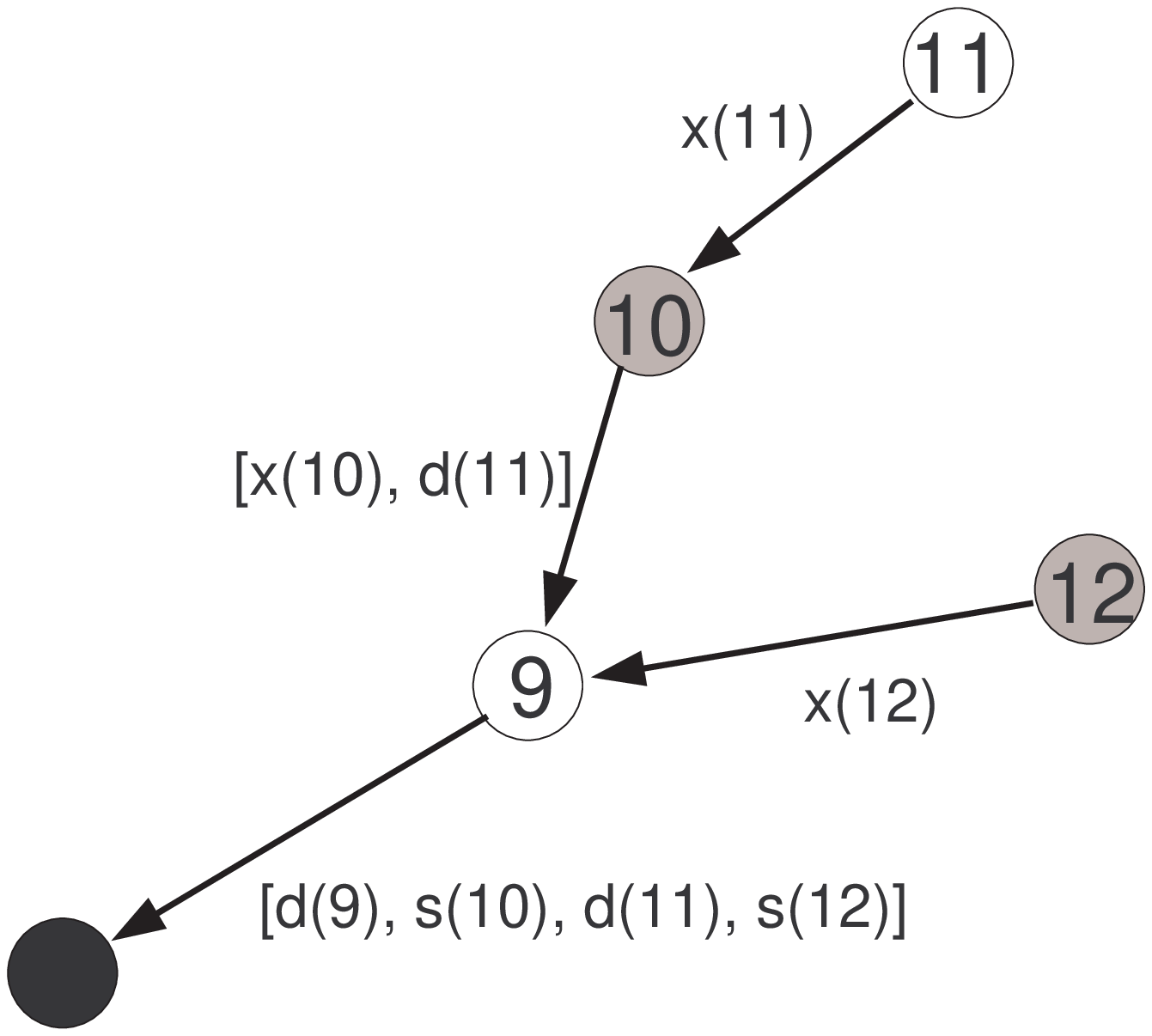}
 \label{fig:Broadcast1_a}}
 %\subfigure[With Broadcasts]{\includegraphics[width=5cm,height=5cm]{Broadcast1_b.eps}
 \subfigure[With Broadcasts]{\includegraphics[width=4cm,height=4cm]{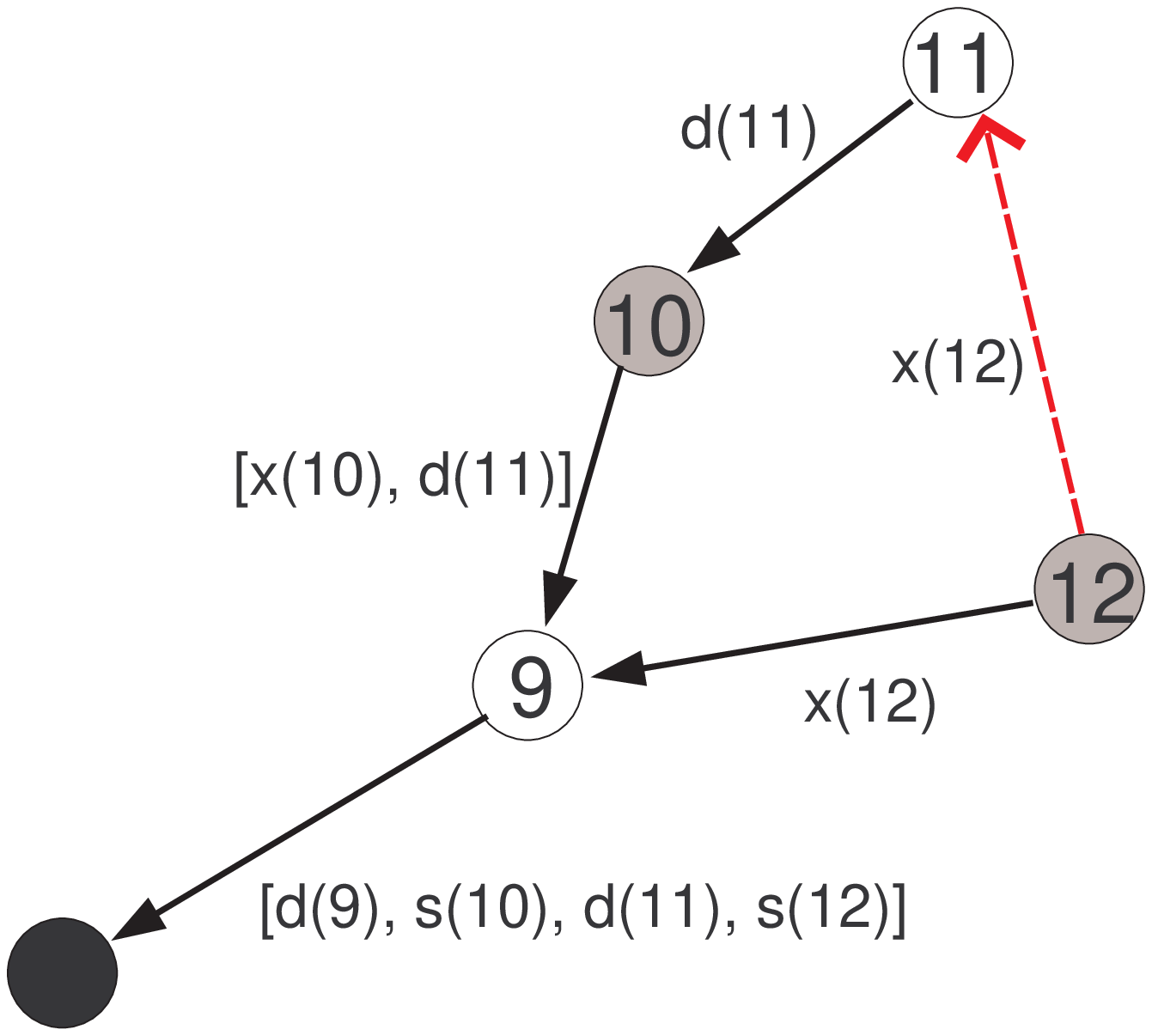}
 \label{fig:Broadcast1_b}}}
 \caption{{\footnotesize No broadcasts are used in (a), so 
 node 11 consumes more resources when transmitting raw data $x(11)$. Broadcasts are used in (b), so 
 node 11 consumes less resources when transmitting detail $d(11)$.}}
 \label{fig:Broadcast1}
\end{figure}

Note that, when broadcast data is used, the decorrelation achieved at odd nodes may still 
be comparable to the 5/3-like transform since the same number of 
neighbors (or more) will be used. 
Moreover, broadcasts are particularly useful for odd nodes $n$ 
that have no children, i.e., $n$ for which $\Cc_n = \emptyset$ but $\Bc_n \cap \Ec \neq \emptyset$. 
If broadcast data is not used when it is available, node $n$ will have to 
transmit $x(n)$ to its parent. Since $x(n)$ requires more bits for encoding than does 
a detail coefficient $d(n)$, $n$ will consume more resources during data transmission. 
By using broadcasts, these odd nodes which have no children can still use 
data overheard from even broadcast neighbors, allowing them to avoid transmitting raw data 
to their parents. This is illustrated in Fig.~\ref{fig:Broadcast1}, where 
node 11 has no children but overhears data from node 12. 
The example in Fig.~\ref{fig:Broadcast1_a} will consume more resources at node 
11 than will the example in Fig.~\ref{fig:Broadcast1_b}. 

\section{Experimental Results}
\label{sec:results}

This section presents experimental results that compare the transforms proposed here against 
existing methods. Source code used to generate these results can be found on our 
webpage\footnote{http://biron.usc.edu/wiki/index.php/Wavelets\_on\_Trees}.
In particular, we focus on comparing the proposed multi-level Haar-like lifting transforms against 
the multi-level 5/3-like transform from~\cite{Shen1,Shen4}, the T-DPCM scheme in~\cite{Shen5} and raw data gathering.  
%Delayed processing for the Haar-like transforms is also used as described in Section~\ref{sec:uni-separable-haar}. 
%Similarly, delayed processing is also used for the 5/3-like transforms as described in~\cite{Shen4}. 
We consider the application of distributed data 
gathering in WSNs. Performance is measured by total energy consumption.
%Since wireless sensor nodes are severely energy-constrained, it is necessary to 
%minimize the amount of data nodes must transfer to the sink in order to reduce 
%the energy nodes consume in the data gathering process. 
%Since data is often spatially correlated across neighboring nodes, we can apply a 
%unidirectional lifting transform in the network to remove 
%spatial data correlation. This will reduce the amount of data nodes must transmit to the sink, 
%thereby reducing the total energy consumption in the network. 
%

\subsection{Experimental Setup}

For evaluation, we consider simulated data generated from 
a second order AR model. This data consists of 
two $600 \times 600$ 2D processes generated by a second order AR model 
with low and high spatial data correlation, e.g., nodes that are a 
certain distance away have higher inter-node correlation for the 
high correlation data than for the low correlation data. 
More specifically, we use the second order AR filter 
$H(z) = \frac{1}{(1-\rho e^{j \omega_0} z^{-1}) (1-\rho e^{-j \omega_0} z^{-1})}$, 
with $\rho = 0.99$ and $\omega_0 = 99$ (resp. $\omega_0 = 359$) 
to produce data with low (resp. high) spatial correlation. 
The nodes were placed in a $600 \times 600$ grid, with node 
measurements corresponding to the data value from the associated 
position in the grid. Each network used in our simulations is generated 
from a set of random node positions distributed in the 
$600 \times 600$ grid. An SPT is constructed for each set of node 
positions. We consider two types of networks: (i) \emph{variable radio range} 
networks in which each node can have a diferent radio range, and (ii) 
\emph{fixed radio range} networks in which each node has the same radio range. 
In the variable radio range case, the radio range that each node $n$ uses 
for transmission is defined by the distance from $n$ to its parent in the SPT. 
Additional broadcast links induced by the SPT are also 
included, i.e., a broadcast link between node $n$ and $m$ exists if 
$m$ is not a direct neighbor of $n$ in the SPT but is within radio range of $n$.

%GSrev1 Shortened this significantly. Not necessary to spell out details found in Wang and LEACH.
%AOrev1 I guess the question to ask is whether we need more information so that it is clear *how* we use the model of Wang and LEACH. 
%GSrev2 Added it back for now...
%In order to measure energy consumption, we use the cost model for WSN devices proposed in~\cite{Wang,LEACH}. 
In order to measure energy consumption, we use the cost model for WSN devices proposed in~\cite{Wang,LEACH}, 
where the energy consumed in transmitting $k$ bits over a distance $D$ is 
$E_T(k,D) = E_{elec} \cdot k + \varepsilon_{amp} \cdot k \cdot D^2$ Joules and the energy consumed in receiving 
$k$ bits is $E_R(k) = E_{elec} \cdot k$ Joules. The $E_{elec} \cdot k$ terms capture the energy dissipated by the 
radio electronics to process $k$ bits. The $\varepsilon_{amp} \cdot k \cdot D^2$ term captures the 
additional energy for signal amplification needed to ensure reasonable signal power at the receiver. 
%%As in~\cite{Wang,LEACH}, we assume that $E_{elec} = 50 \text{ nJ/bit}$ and $\varepsilon_{amp} = 100 \text{ pJ/bit/m}^2$. 
WSN devices also consume energy when performing computations, but 
these costs are typically very small compared with transmission and reception costs. 
Therefore, we ignore them in our cost computations. 
%For the sake of simplicity, we also assume that $\varepsilon_{amp} \cdot D^2 >> E_{elec}$, e.g., 
%nodes are spaced far enough apart that the energy consumed in transmitting 
%1 bit of data is much higher than the energy consumed by the radio electronics for 
%processing 1 bit of data.
Also note that all data gathering schemes will suffer from channel noise and attenuation, 
so a no-channel-loss comparison is still valid. Thus, we do not consider these effects 
in our experiments.

Comparisons are made with the Haar-like transforms of 
Section~\ref{sec:uni-separable-haar} against 
the 5/3-like transform with delayed processing proposed in~\cite{Shen4} and the 
T-DPCM scheme proposed in~\cite{Shen5}. 
Predictions for each of these transforms are made using the adaptive prediction 
filter design in~\cite{Shen5}. Updates are made using the 
``orthogonalizing'' update filter design in~\cite{Shen6}. 
In each epoch, we assume that each node transmits $M = 50$ measurements taken at $M$ different times. 
Also, each raw measurement is represented using $B_r = 12$ bits. 
%We use integer to integer operations~\cite{calderbank} 
%in all of the transform computations (i.e., 
%$d(n) = x(n) + \lfloor \sum_{i \in \Nc_n} \pv_n(i) x(i) \rfloor$ 
%and $s(n) = x(n) + \lfloor \sum_{j \in \Nc_n} \uv_n(j) d(j) \rfloor$) to allow 
%for lossless representations. 
We assume each odd node encodes $M$ detail coefficients together 
with an adaptive arithmetic coder. 
%using an adaptive arithmetic coder~\cite{Langdon}. 
Smooth coefficients are treated like raw data, i.e., each one uses $B_r$ bits. 
Since we only seek to compare the performance of spatial transforms, 
we do not consider any temporal processing. 

\subsection{Simulation Results}

\begin{figure}[htb]
 %\centerline{\subfigure[Variable Radio Range]{\includegraphics[width=7cm,height=7cm]{LosslessCost-a.eps}
 \centerline{\subfigure[Variable Radio Range]{\includegraphics[width=6.5cm,height=6.5cm]{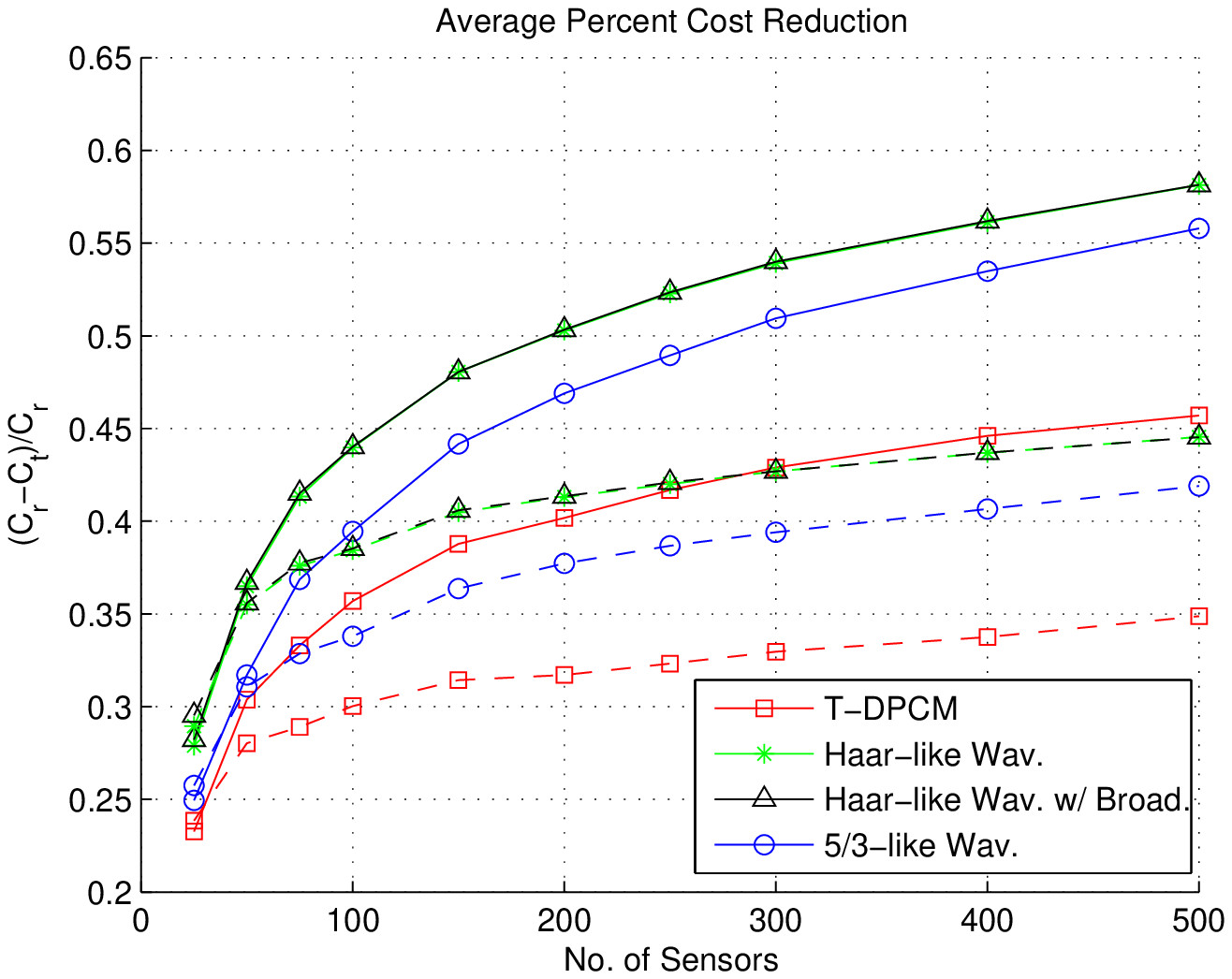}
 \label{fig:LosslessCost-a}}
 %\subfigure[Fixed Radio Range]{\includegraphics[width=7cm,height=7cm]{LosslessCost-b.eps}
 \subfigure[Fixed Radio Range]{\includegraphics[width=6.5cm,height=6.5cm]{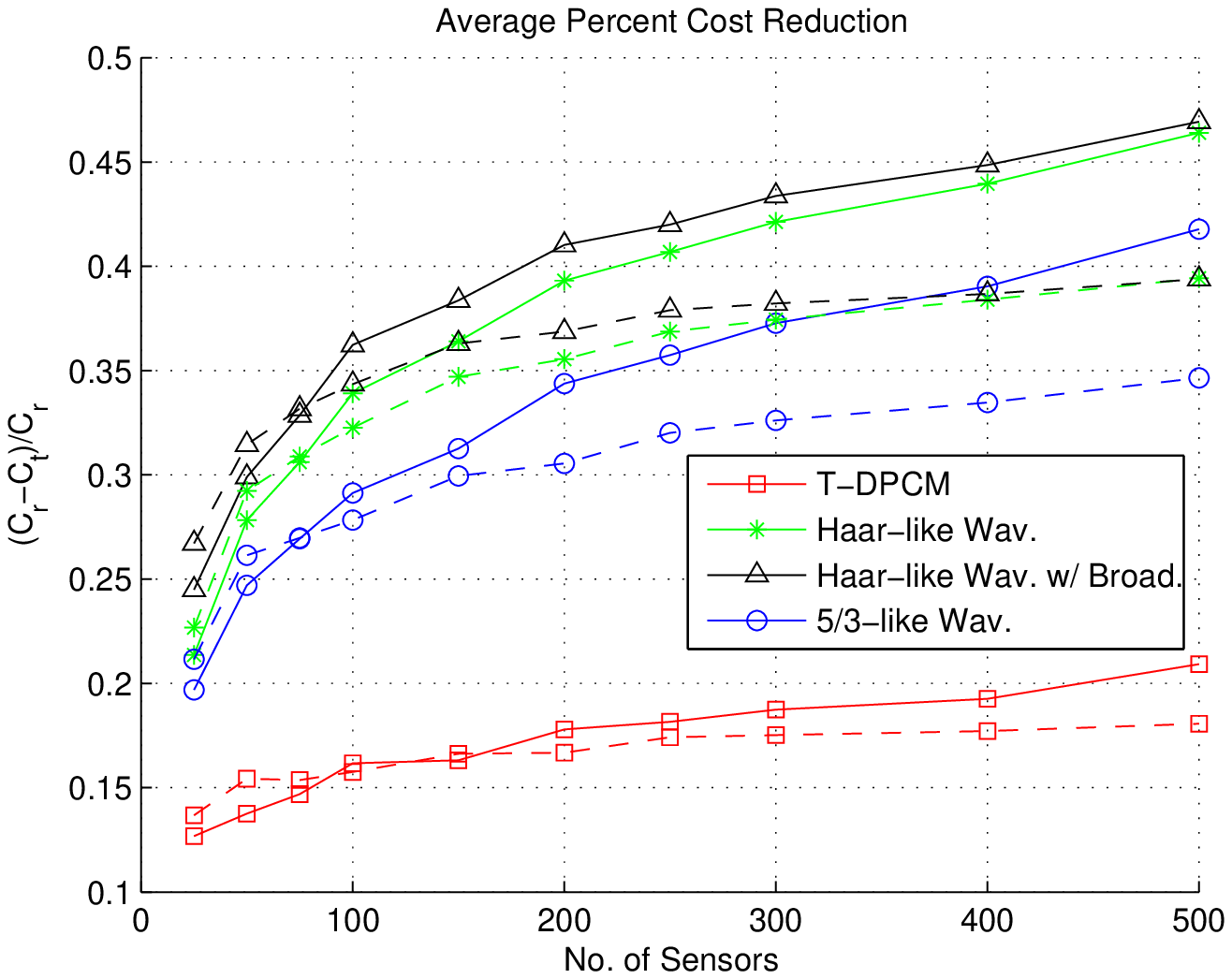}
 \label{fig:LosslessCost-b}}}
 \caption{{\footnotesize Average percent cost reduction ($\frac{C_r-C_t}{C_r}$). Solid and dashed lines correspond to 
 high and low spatial data correlation, respectively. Best performance achieved by Haar-like transforms, 
 followed by 5/3-like transform and T-DPCM. High correlation data also gives greater cost 
 reduction than low correlation data.}}
 \label{fig:lossless}
\end{figure}

In the case of lossless compression, the average cost reduction ratios taken over multiple 
uniformly distributed networks are shown in Fig.~\ref{fig:lossless} for high and low data correlation. 
These are expressed as the average of multiple values of $(C_r-C_t)/C_r$, 
where $C_t$ is the cost for joint routing and transform and $C_r$ is the cost for raw data forwarding. 
%GSrev1 Removed rate reduction ratios to make room for experiments showing fixed radio range cases
%The average rate reduction ratios (i.e., $c_{avg} = \frac{\sum_n B_n/N}{B_r}$) are shown in Fig.~\ref{fig:LosslessRRR} 
%for the Haar-like and 5/3-like transforms for high and low data correlation. 
Results for variable radio ranges (each node has different radio range) are shown in 
Fig.~\ref{fig:LosslessCost-a}. Results for fixed radio ranges (each node has the 
same radio range) are given in Fig.~\ref{fig:LosslessCost-b}. 
T-DPCM does the worst overall. The 5/3-like transform provides significant improvement over the 
simple T-DPCM scheme. 
The Haar-like transforms have the highest average cost reduction ratio, 
or equivalently, the lowest average cost. 
Moreover, we note that broadcast is not very helpful (on average) when nodes have variable 
radio ranges (Fig.~\ref{fig:LosslessCost-a}), but there is a significant gain when 
nodes use a fixed radio range (Fig.~\ref{fig:LosslessCost-b}). 
This is mainly because, in the fixed radio range case, (i) there are many more opportunities for using broadcast data 
and (ii) each node has more broadcast neighbors. 
%Thus, broadcast is likely to be most useful when nodes are configured with a fixed radio range. 

%\begin{figure}[htb]
% \centerline{\subfigure[Cost Reduction Ratios]{\includegraphics[width=7cm,height=7cm]{LosslessCost.eps}
% \label{fig:LosslessCost}}
% \subfigure[Rate Reduction Ratios]{\includegraphics[width=7cm,height=7cm]{LosslessRRR.eps}
% \label{fig:LosslessRRR}}}
% \caption{{\footnotesize Cost and rate reduction ratios. Solid and dashed lines correspond to 
% high and low spatial data correlation, respectively. Best performance achieved by Haar-like transform, 
% followed by 5/3-like transform and T-DPCM. High correlation data also gives greater cost 
% reduction than low correlation data.}}
% \label{fig:lossless}
%\end{figure}
%

Note that the amount of raw data forwarding needed to compute the Haar-like transform 
is significantly reduced compared with the 5/3-like transform. 
Therefore, the Haar-like transform will do better than the 5/3-like transform 
in terms of transport costs. Granted, the 5/3-like transform will use data from more neighbors 
for processing, so the decorrelation given by the 5/3-like transform will be 
greater than that given by the Haar-like transform. 
%GSrev1 Again, shortened this discussion a bit
However, in our experiments the average reduction in rate that the 5/3-like 
transform provides over the Haar-like transform is rather small. 
%However, by examining the rate reduction ratios in Fig.~\ref{fig:LosslessRRR}, 
%it is clear that the additional reduction in rate provided by the 5/3-like transform is very small. 
%This small reduction in the total bit rate with respect to the Haar-like transform is not likely to offset the 
%larger amount of raw data forwarding. Therefore, for this type of data, it is almost never worthwhile to do an 
%additional step of raw data forwarding to compute a detail coefficient. 
%Using broadcasts in the Haar-like transform provides an (average) additional reduction of up to 2\% percent. 
The Haar-like transform with broadcast also provides additional cost reduction over the Haar-like transform without 
broadcasts since less raw data forwarding is needed on average. 
Moreover, the amount of cost reduction achievable is higher for the high 
correlation data than for the low correlation data.
%GSrev1 Reworded this since we no longer show rate reduction ratios
%Moreover, the amount of cost and rate reduction achievable is higher for the high 
%correlation data than for the low correlation data, which is what we would expect.

\begin{figure}[htb]
 \centerline{\subfigure[Variable RR Network]{\includegraphics[width=6cm,height=6cm]{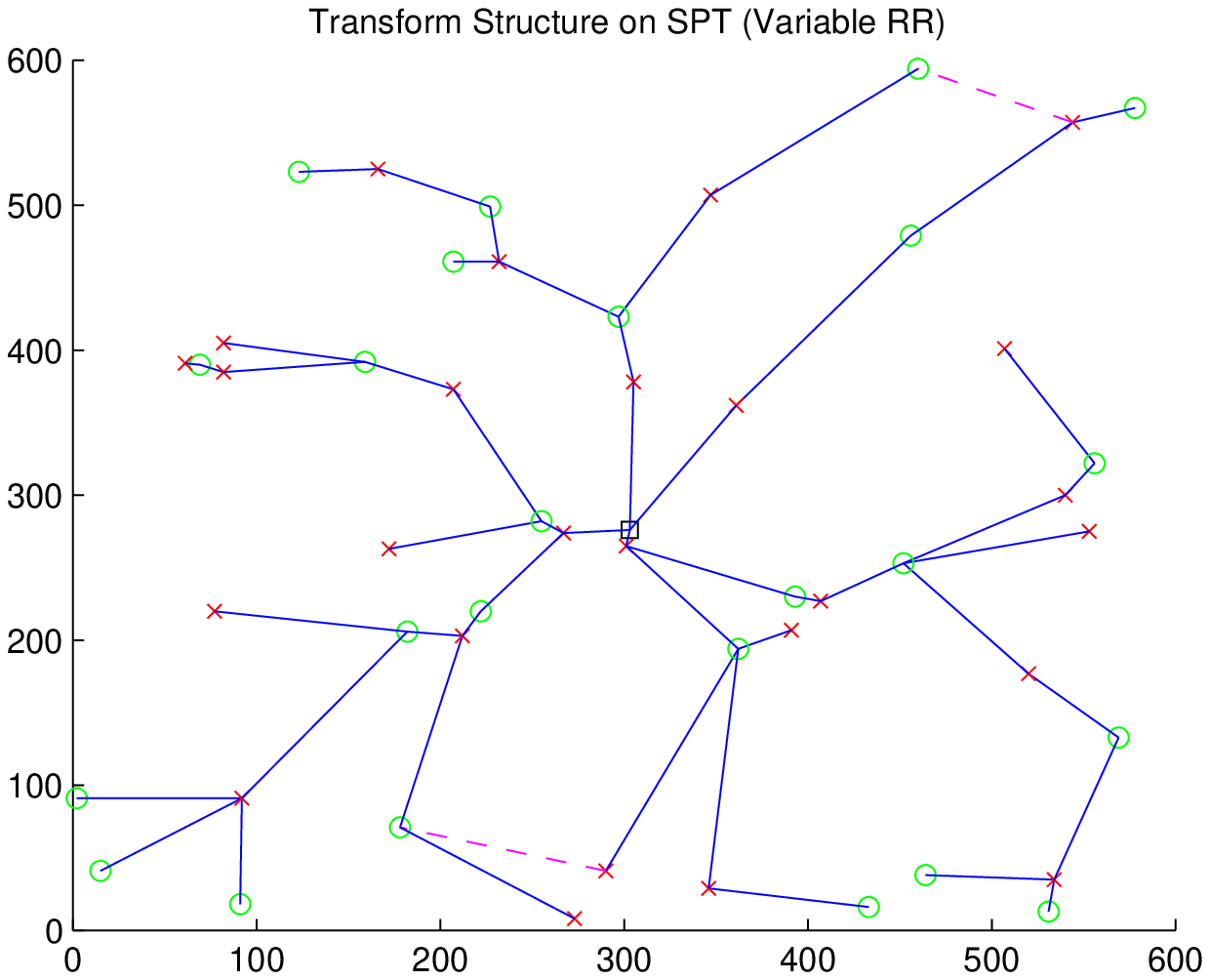}
 \label{fig:lossyNet}}
 \subfigure[Cost-Distortion Curves (Variable RR)]{\includegraphics[width=6cm,height=6cm]{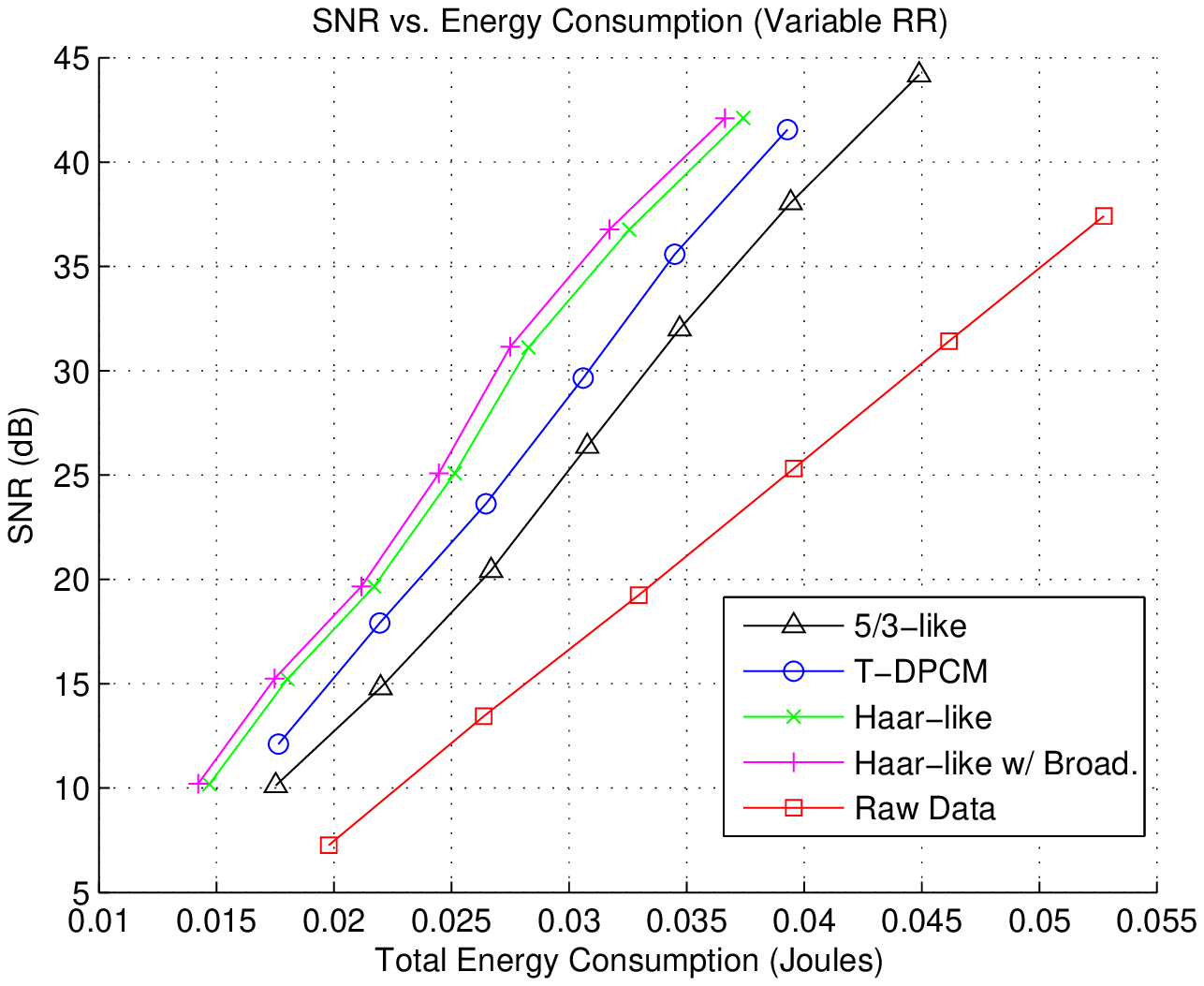}
 \label{fig:lossyCD}}}
 \centerline{\subfigure[Fixed RR Network]{\includegraphics[width=6cm,height=6cm]{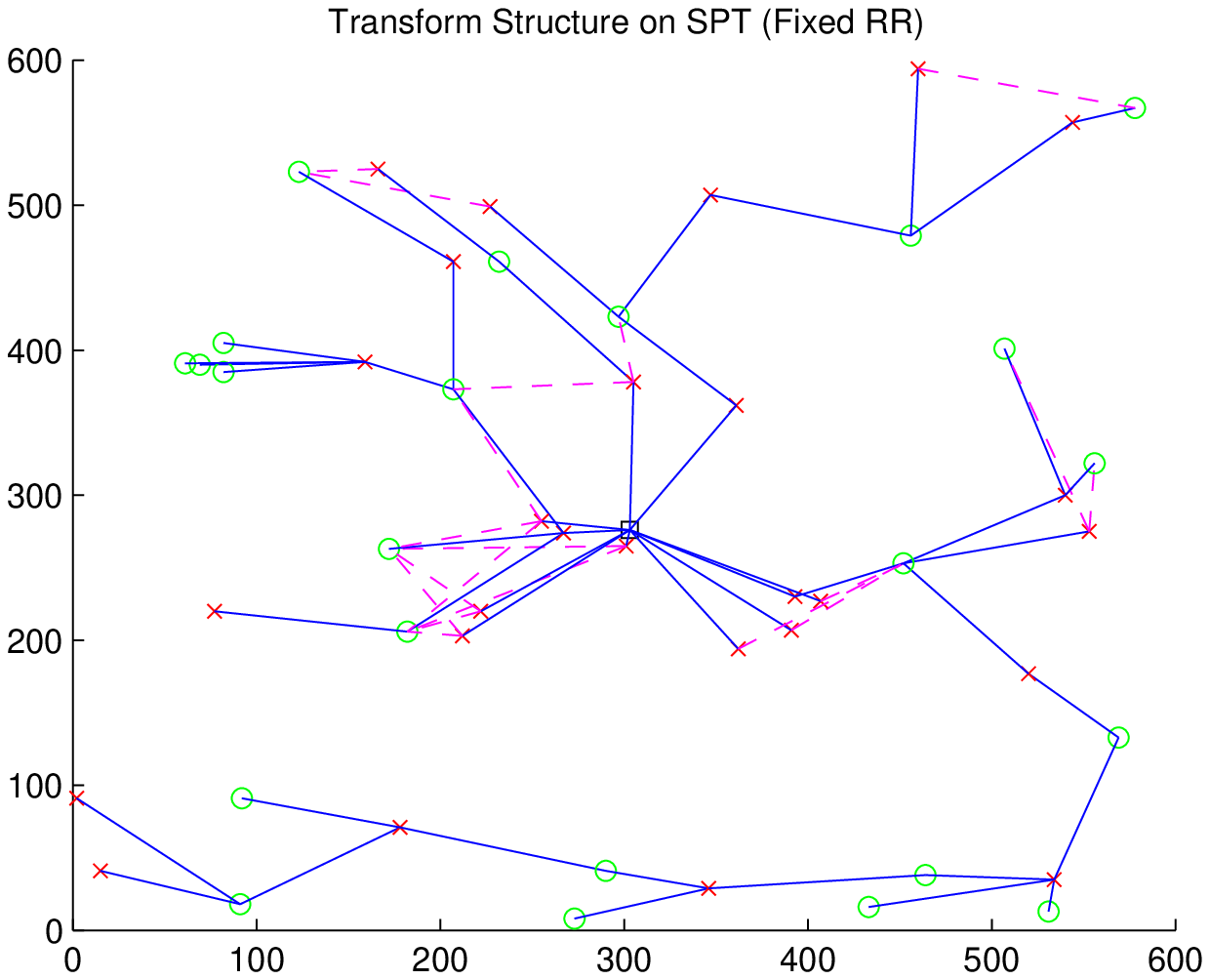}
 \label{fig:lossyNet_Fixed}}
 \subfigure[Cost-Distortion Curves (Fixed RR)]{\includegraphics[width=6cm,height=6cm]{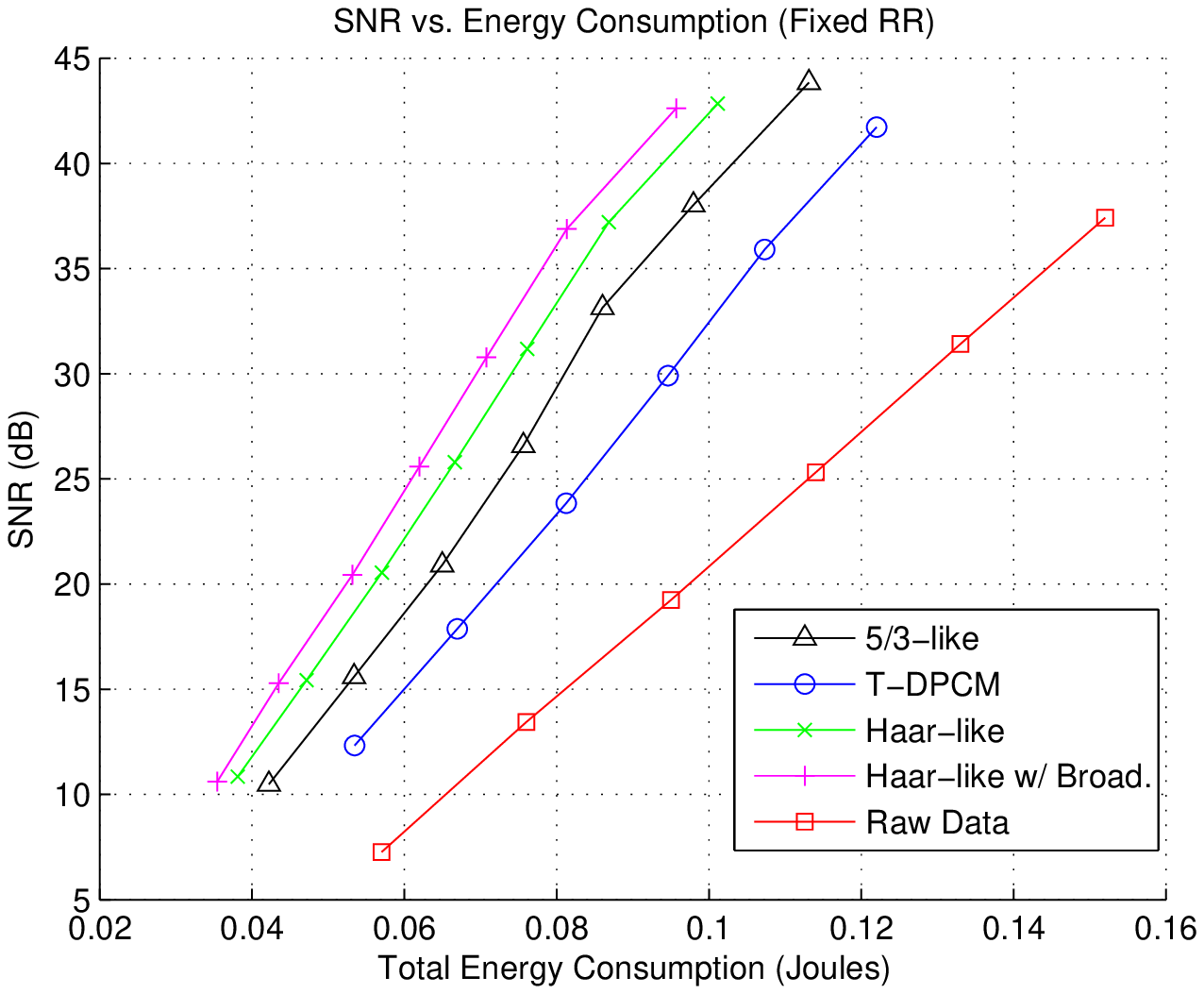}
 \label{fig:lossyCD_Fixed}}}
% \caption{{\footnotesize Sample network with corresponding Cost-Distortion curves. Green circles represent 
% even nodes, red x's represent odd nodes.}}
  \caption{{\footnotesize Sample networks with corresponding Cost-Distortion curves. In (a) and (c), blue lines denote 
  forwarding links, dashed magenta lines are broadcast links, green circles are even nodes, red x's 
  are odd nodes, and the black center node is the sink.}}
 \label{fig:lossy}
\end{figure}

Lossy coding is also possible and can provide even greater cost reductions while introducing some 
reconstruction error. In this case, we quantize transform coefficients with a dead-zone uniform 
scalar quantizer. 
%scalar quantizer~\cite{Taubman}. 
Performance is measured by the trade-off between total cost and distortion in the reconstructed data, which 
we express as the signal to quantization noise ratio (SNR). Sample 50 node networks are shown 
in Figs.~\ref{fig:lossyNet} and~\ref{fig:lossyNet_Fixed} and, 
in the case of high correlation data, the corresponding performance 
curves are shown in Figs.~\ref{fig:lossyCD} and~\ref{fig:lossyCD_Fixed}. 
The Haar-like transforms do the best among all transforms. 
%GSrev1 This sentence is repetitive
%This is reasonable since it has the lowest amount of raw data forwarding cost among these transforms, 
%hence the total cost will also be lower. 

%\begin{figure}[htb]
% \centerline{\subfigure[Sample Network]{\includegraphics[width=7cm,height=7cm]{lossyNet.eps}
% \label{fig:lossyNet}}
% \subfigure[Cost-Distortion Curves]{\includegraphics[width=7cm,height=7cm]{lossyCD.eps}
% \label{fig:lossyCD}}}
%% \caption{{\footnotesize Sample network with corresponding Cost-Distortion curves. Green circles represent 
%% even nodes, red x's represent odd nodes.}}
%  \caption{{\footnotesize Sample network with corresponding Cost-Distortion curves. In (a), blue lines denote 
%  forwarding links, dashed magenta lines denote broadcast links, green circles denote even nodes, red x's 
%  denote odd nodes, and the black center node is the sink.}}
% \label{fig:lossy}
%\end{figure}
%
%
%\begin{figure}[htb]
% \centerline{\subfigure[Variable RR Network]{\includegraphics[width=4cm,height=4cm]{lossyNet.eps}
% \label{fig:lossyNet}}
% \subfigure[Variable RR Performance]{\includegraphics[width=4cm,height=4cm]{lossyCD.eps}
% \label{fig:lossyCD}}
% \subfigure[Fixed RR Network]{\includegraphics[width=4cm,height=4cm]{lossyNet_Fixed.eps}
% \label{fig:lossyNet_Fixed}}
% \subfigure[Fixed RR Performance]{\includegraphics[width=4cm,height=4cm]{lossyCD_Fixed.eps}
% \label{fig:lossyCD_Fixed}}}
%% \caption{{\footnotesize Sample network with corresponding Cost-Distortion curves. Green circles represent 
%% even nodes, red x's represent odd nodes.}}
%  \caption{{\footnotesize Sample networks with corresponding Cost-Distortion curves. In (a) and (c), blue lines denote 
%  forwarding links, dashed magenta lines are broadcast links, green circles are even nodes, red x's 
%  are odd nodes, and the black center node is the sink.}}
% \label{fig:lossy}
%\end{figure}

%GSrev5 Re-wrote this paragraph to reflect the new experiments
When using broadcasts with the Haar-like transform, 
there is an additional 1 dB (resp. 2.5 dB) gain in SNR for the variable (resp. fixed) radio range 
network at a fixed cost, i.e., by using broadcasts we can increase the quality in the 
reconstructed data for a fixed communication cost. Thus, for these networks, using broadcast is quite 
helpful. Also note that there only 2 broadcast links used in the transform for the variable radio 
range network (Fig.~\ref{fig:lossyNet}), whereas there are over 10 broadcast links used in the 
fixed radio range network (Fig.~\ref{fig:lossyNet_Fixed}).  
Thus, broadcast provides even greater gains for the fixed radio range network (2.5 dB versus 1dB) 
since there are more broadcast links. More generally, broadcast should provide 
more gains in networks where many broadcast opportunities are available.
%, particularly because there is an odd node (at the bottom) which has no children and its parent 
%is also very far away. If that broadcast link is not used, this node must transmit 
%raw data over a very long distance and this will be rather costly. Instead, by using broadcast it can 
%locally compute detail coefficients, which can then be encoded using very few bits. 
%This will significantly reduce the transmission cost for this particular odd node. 
%If a network were to have many more such odd 
%nodes, then we would expect the benefits of using broadcast to be even greater. This is more likely to happen 
%in networks where nodes use fixed radio ranges, thus, we should expect more gains from broadcasts in those networks.

Also note that in this particular network for the variable radio range case, T-DPCM actually does better than the 5/3-like transform. 
Note that in T-DPCM, only the 
leaf nodes forward raw data to the sink; so if there are only a few leaf nodes, the raw data forwarding cost 
for T-DPCM may not be very high compared with the raw data forwarding cost for the 
5/3-like transform. In this particular network, only 19 of the 50 nodes are leaves in the tree. 
Therefore, the raw data forwarding cost for T-DPCM in this case is lower than that for the 5/3-like transform. 
%This is the main reason why T-DPCM does better than the 5/3-like wavelet for this network. 
However, on average the raw data forwarding cost for T-DPCM will be very high (see Fig.~\ref{fig:lossless}), 
leading to higher total cost on average as compared with the 5/3-like transform. 

\section{Conclusions}
\label{sec:conclusions}

A general class of en-route in-network (or unidirectional) transforms has been proposed along with 
a set of conditions for their invertibility. This covers a wide range of existing unidirectional 
transforms and has also led to new transform designs which outperform the existing transforms 
in the context of data gathering in wireless sensor networks. In particular, we have used the 
proposed framework to provide a general class of invertible unidirectional wavelet transforms 
constructed using lifting. These general wavelet transforms can also take into account 
broadcast data without affecting invertibility. A unidirectional Haar-like transform was also proposed which 
significantly reduces the amount of raw data transmissions that nodes need to make. Since 
raw data requires many more bits than encoded transform coefficients, this leads to a significant 
reduction in the total cost. Moreover, our proposed framework allows us to easily incorporate broadcasts 
into the Haar-like transforms without affecting invertibility. This use of broadcast data provides 
further performance improvements for certain networks. 

\bibliographystyle{IEEEbib}
%{\small 
\bibliography{WSN_Journal_Paper}
%}

%\begin{IEEEbiographynophoto}{Godwin Shen}
%(S'07) received the B.S. degree from California State Polytechnic University, Pomona, in 
%2005 and the M.S. degree from University of Southern California (USC), Los Angeles, in 2007, where 
%he is currently pursuing his PhD degree.
%\end{IEEEbiographynophoto}
%
%\begin{IEEEbiographynophoto}{Antonio Ortega}
%Biography text here.
%\end{IEEEbiographynophoto}

\end{document}

%% file: macros.tex
\long\def\comment#1{}

\newfont{\bbb}{msbm10 scaled 700}

\newfont{\bb}{msbm10 scaled 1100}

\newcommand{\av}{{\bf a}}

\newcommand{\pv}{{\bf p}}

\newcommand{\uv}{{\bf u}}

\newcommand{\xv}{{\bf x}}
\newcommand{\yv}{{\bf y}}

\newcommand{\Am}{{\bf A}}
\newcommand{\Bm}{{\bf B}}
\newcommand{\Cm}{{\bf C}}

\newcommand{\Gm}{{\bf G}}
\newcommand{\Hm}{{\bf H}}
\newcommand{\Id}{{\bf I}}

\newcommand{\Mm}{{\bf M}}

\newcommand{\Pm}{{\bf P}}

\newcommand{\Um}{{\bf U}}

\newcommand{\Ac}{{\cal A}}
\newcommand{\Bc}{{\cal B}}
\newcommand{\Cc}{{\cal C}}
\newcommand{\Dc}{{\cal D}}
\newcommand{\Ec}{{\cal E}}

\newcommand{\Ic}{{\cal I}}

\newcommand{\Nc}{{\cal N}}
\newcommand{\Oc}{{\cal O}}

\newcommand{\Rc}{{\cal R}}

\renewcommand{\det}{{\hbox{det}}}